\documentclass[11pt]{article} % font size
\usepackage[english]{babel}
\usepackage[a4paper,margin=1in]{geometry} % page setup
\usepackage{authblk} % author affiliations
\usepackage[colorlinks=true, linkcolor=black, citecolor=black, urlcolor=blue, filecolor=blue,breaklinks=true]{hyperref} % Hyperlinks, needed for biblatex, other option: allcolors=blue
\usepackage{setspace} % linespacing
\usepackage{lineno} % line numbering

\usepackage{csquotes} % Recommended for biblatex, must load after lineno or will get a warning
\usepackage{ragged2e}% for left aligning
\usepackage{array} % used to soft wrap text in Table S3

\usepackage[scaled=0.9]{helvet}

\usepackage[defaultmathsizes]{mathastext}
\usepackage{amsmath}

\usepackage[capitalise, noabbrev, nameinlink]{cleveref}
\usepackage[labelsep=period]{caption} % change separator to period instead of colon
\crefdefaultlabelformat{#2#1#3}
\Crefname{figure}{Figure}{Figures}
\Crefname{table}{Table}{Tables}
\Crefname{section}{\textbf{Section}}{\textbf{Sections}}
\usepackage{graphicx}
\graphicspath{{./main_figs/}{./supplement/suppl_figs/}} % paths for figs
\DeclareGraphicsExtensions{.pdf,.jpeg,.JPG,.png,.PNG, .eps, .tiff}
\usepackage{subcaption} % for subcaptions (panels), and add parentheses
\DeclareCaptionLabelFormat{bold}{{(#2)}} % bold subpanel letter in caption
\newcommand{\labelphantom}[1]{%  To make subpanel references easier from https://tex.stackexchange.com/a/255790/121424 
  \parbox{0pt}{\phantomsubcaption\label{#1}}%
}
\usepackage{pgffor} % for function to make figures with subpanels
\usepackage{alphalph} % for function to make figures with subpanels
\usepackage[labelfont=bf, textfont=bf, singlelinecheck=off, textfont=footnotesize]{caption} % boldness of captions
\usepackage{booktabs}
\usepackage{multirow}
\usepackage{rotating}

\newcommand{\generateFigSubpanels}[6][0]{ % default rotation = 0
     \expandafter\newcommand\csname#2\endcsname{ %https://tex.stackexchange.com/questions/65780/macro-defining-macro/65781#65781
      \begin{figure}[htbp]
        \foreach [count=\i] \x in #6{%
            \labelphantom{fig:#2:\AlphAlph{\i}}
        }
        \centering
        \includegraphics[width=#4\linewidth,angle=#1]{#3}
        \caption{\textbf{\color{sectioncolor}\normalsize #5}}\label{fig:#2}
        \footnotesize
        \justifying
        \foreach [count=\i] \x in #6{%
            \noindent\subref*{fig:#2:\AlphAlph{\i}}~\x\space\\ % added * after "subref" to disable links and "//" at end to make each subpanel caption a new line
        }
      \end{figure}
    }
}

\newcommand{\generateFig}[6][0]{ % default rotation = 0
     \expandafter\newcommand\csname#2\endcsname{%https://tex.stackexchange.com/questions/65780/macro-defining-macro/65781#65781
      \begin{figure}[htbp]
        \centering
        \includegraphics[width=#4\linewidth,angle=#1]{#3}
        \caption{\textbf{\color{sectioncolor}\normalsize #5}\label{fig:#2}
        \footnotesize
        \normalfont
        #6
        }
        
      \end{figure}
    }
}

\newcommand{\generateSidewaysFigSubpanels}[6][0]{% default rotation=0
     \expandafter\newcommand\csname#1\endcsname{ %https://tex.stackexchange.com/questions/65780/macro-defining-macro/65781#65781
      \begin{figure}[htbp]
        \foreach [count=\i] \x in #5{%
            \labelphantom{fig:#2:\AlphAlph{\i}}
        }
        \centering
        \includegraphics[width=#3\linewidth,angle=#1]{#3}
        \caption{\textbf{\color{sectioncolor}\normalsize #5}}\label{fig:#2}
        \footnotesize
        \justifying
        \foreach [count=\i] \x in #6{%
            \noindent\subref*{fig:#2:\AlphAlph{\i}}~\x\space\\ % added * after "subref" to disable links and "//" at end to make each subpanel caption a new line
        }
      \end{figure}
    }
}

\newcommand{\generateSidewaysFig}[6][0]{%  default rotation=0
     \expandafter\newcommand\csname#2\endcsname{ %https://tex.stackexchange.com/questions/65780/macro-defining-macro/65781#65781
      \begin{sidewaysfigure}[htbp]
        \centering
        \includegraphics[width=#4\linewidth,angle=#1]{#3}
        \caption{\textbf{\color{sectioncolor}\normalsize #5}\label{fig:#2}
        \footnotesize
        \normalfont
        #6
        }
        
      \end{sidewaysfigure}
    }
}

\newcommand{\generateTab}[6][]{%  Note that rotation does not work here
     \expandafter\newcommand\csname#2\endcsname{ \begin{table}[ht]
          \centering
                \resizebox{#4\textwidth}{!}{
                \input{#3}
            }
            \caption{\textbf{\normalsize #5}\label{tab:#2}
            \footnotesize
            \normalfont
            #6
            }
            
        \end{table}
    }
}

\newcommand{\generateSidewaysTab}[6][]{%  Note that rotation does not work here
     \expandafter\newcommand\csname#2\endcsname{ \begin{sidewaystable}[ht]
          \centering
                \resizebox{#4\textwidth}{!}{
                \input{#3}
            }
            \caption{\textbf{\normalsize #5}\label{tab:#2}
            \footnotesize
            \normalfont
            #6
            }
            
        \end{sidewaystable}
    }
}

\usepackage[dvipsnames]{xcolor} % can remove, for coloring the tips
\usepackage{changepage} % can remove, for width on the tips
\usepackage{enumitem} % can remove, for width on the tips

\usepackage{xurl}
\usepackage[backend=biber, citestyle=numeric-comp, bibstyle=authoryear, sorting=none, natbib=true, url=false, minbibnames=10, maxbibnames=10, uniquename=false, uniquelist=false, giveninits=true]{biblatex}

\defbibenvironment{bibliography}
  {\list
    {\printtext[labelnumberwidth]{%
       \printfield{labelprefix}%
       \printfield{labelnumber}}}
    {\setlength{\labelwidth}{\labelnumberwidth}%
     \setlength{\leftmargin}{\labelwidth}%
     \setlength{\labelsep}{\biblabelsep}%
     \addtolength{\leftmargin}{\labelsep}%
     \setlength{\itemsep}{\bibitemsep}%
     \setlength{\parsep}{\bibparsep}}%
     }
  {\endlist}
  {\item}

\DeclareFieldFormat{labelnumberwidth}{#1.}

\DeclareNameAlias{author}{family-given}

\AtEveryBibitem{\clearfield{month}}
\AtEveryBibitem{\clearfield{number}}
\AtEveryBibitem{\clearfield{issn}}

\DeclareFieldFormat[article]{title}{#1}

\DeclareFieldFormat{journaltitle}{#1\isdot}

\renewbibmacro{in:}{}

\renewbibmacro*{volume+number+eid}{%
  \printfield{volume}%
}
\DeclareFieldFormat[article]{volume}{\mkbibitalic{#1}}

\DeclareFieldFormat{doi}{\url{https://doi.org/#1}}

\DefineBibliographyStrings{english}{%
  page             = {},
  pages            = {},
} 

\usepackage{xcolor}
\usepackage{caption}
\usepackage{dsfont}
\definecolor{sectioncolor}{HTML}{9c4847}
\definecolor{black}{HTML}{000000}
\definecolor{linkcolor}{HTML}{008ebe}

\usepackage{titlesec}
\titleformat{\section}
{\color{sectioncolor}\normalfont\Large\bfseries}
{\thesection}{1em}{}

\titleformat{\subsection}
{\color{sectioncolor}\normalfont\large\bfseries}
{\thesubsection}{1em}{}

\DeclareCaptionFont{figuretitlecolor}{\color{sectioncolor}}
\DeclareCaptionFont{figurecaptioncolor}{\color{black}}
\DeclareCaptionFont{tabletitlecolor}{\color{sectioncolor}}
\DeclareCaptionFont{tablecaptioncolor}{\color{black}\footnotesize}
\hypersetup{
    colorlinks=true,
    linkcolor=linkcolor,
    urlcolor=linkcolor,
    citecolor=linkcolor
}

\makeatletter
\renewcommand\maketitle{\par
  \begingroup
    \flushleft
    \Large{\@title}\par
  \endgroup
  \vspace{0.5em}
  \begin{flushleft}
    \@author
  \end{flushleft}
}
\makeatother

\renewenvironment{abstract}{
  \begin{flushleft}
    \textbf{\abstractname}
  \end{flushleft}
  \vspace{-1.5em}
  \par
  \noindent\justify
  \parfillskip=0pt
}{
  \par 
}

\title{Targeted incentives for social tipping in heterogeneous networked populations}

\author[1,*]{Dhruv Mittal}
\author[2]{Fátima González-Novo López}
\author[3,4]{Sara M. Constantino}
\author[5]{Shaul Shalvi}
\author[6]{Xiaojie Chen}
\author[1,7,*]{Vítor V. Vasconcelos}

\affil[1]{Computational Science Lab, Informatics Institute, University of Amsterdam, 1098 XH Amsterdam, The Netherlands}
\affil[2]{Graduate School of Informatics, University of Amsterdam, 1098 XH Amsterdam, The Netherlands}
\affil[3]{Doerr School of Sustainability, Stanford University, Stanford, CA 94305, USA}
\affil[4]{School of Public and International Affairs, Princeton University, Princeton, NJ 08544, USA}
\affil[5]{Center for Research in Experimental Economics and political Decision making (CREED), Amsterdam School of Economics, University of Amsterdam, Amsterdam, The Netherlands}
\affil[6]{School of Mathematical Sciences, University of Electronic Science and Technology of China, Chengdu 611731, People’s Republic of China}
\affil[7]{POLDER center, Institute for Advanced Study, University of Amsterdam, 1012 GC Amsterdam, The Netherlands}

\affil[*]{Correspondence: \href{mailto:d.mittal@uva.nl}{d.mittal@uva.nl} and  \href{mailto:v.v.vasconcelos.uva.nl}{v.v.vasconcelos.uva.nl} }
\affil[ ]{ } % for some blank space
\date{} % If you want a date fill this in (e.g. \date{February 2022})

\newcommand{\makeAbstract}{
\renewcommand{\abstractname}{\color{sectioncolor}ABSTRACT}
\begin{abstract}
\noindent \color{linkcolor}Many societal challenges, such as climate change or disease outbreaks, require coordinated behavioral changes. For many behaviors, the tendency of individuals to adhere to social norms can reinforce the status quo. However, these same social processes can also result in rapid, self-reinforcing change. Interventions may be strategically targeted to initiate endogenous social change processes, often referred to as social tipping. While recent research has considered how the size and targeting of such interventions impact their effectiveness at bringing about change, they tend to overlook constraints faced by policymakers, including the cost, speed, and distributional consequences of interventions. To address this complexity, we introduce a game-theoretic framework that includes heterogeneous agents and networks of local influence. We implement various targeting heuristics based on information about individual preferences and commonly used local network properties to identify individuals to incentivize. Analytical and simulation results suggest that there is a trade-off between preventing backsliding among targeted individuals and promoting change among non-targeted individuals. Thus, where the change is initiated in the population and the direction in which it propagates is essential to the effectiveness of interventions. We identify cost-optimal strategies under different scenarios, such as varying levels of resistance to change, preference heterogeneity, and homophily. These results provide insights that can be experimentally tested and help policymakers to better direct incentives.\\
\end{abstract}
}

\begin{document}
    \setstretch{1.15} %\onehalfspace
    %\linenumbers
    %\input{REMOVEME_tips}

    \maketitle
    \makeAbstract
    \clearpage
    \section{INTRODUCTION}
%\introTips % Delete when no longer needed

% ADD emphasis on the difference between - back-sliding - trade off of localizing . 

Societies around the world face urgent challenges that require large-scale coordinated changes in behaviors, technologies, and infrastructures \cite{Creutzig2022}. However, real-world transitions often stall due to entrenched social norms \cite{nyborg2016social, gavrilets2020dynamics}, biased perceptions \cite{santos2021biased}, low risk-awareness or high uncertainty \cite{santos2011risk,vasconcelos2015cooperation,weber2006experience}, and misaligned incentives \cite{mittal2024anti}. For example, addressing climate change will require households to reduce their meat consumption and to fly or drive less. However, these behaviors are reinforced by prevailing social norms and can be difficult to change \cite{efferson2020promise}---and this is true for many collective action problems, where behaviors are socially interdependent. In such cases, interventions that increase the incentives of households to switch to desirable behaviors may be necessary \cite{dietz2009household, otto2020social}. At the same time, interventions aimed at shifting behaviors away from the status quo can be costly, politically challenging to implement and, once implemented, may face backlash or have limited effects. In response to these challenges, researchers have suggested that even targeted interventions may be sufficient to initiate endogenous social change \cite{nyborg2016social}, potentially reducing some of these concerns. However, while the promise that circumscribed interventions can tip societies towards new norms (i.e., social tipping) has gained traction in the climate solutions literature, there is limited research on how such interventions should be designed given constraints faced by policymakers \cite{pizziol2024niches, constantino2022scaling}. Here, we address this gap by considering how the size and targeting of circumscribed interventions \cite{efferson2020promise, ehret2022group}, as well as the social network structure \cite{banerjee2013diffusion,vasconcelos2021segregation}, impact the extent, speed, cost, and durability of behavior change. 

Thus, a key open question is how to design interventions to strategically and cost-effectively "kickstart" endogenous behavioral change processes \cite{pizziol2024niches, constantino2022scaling}. %This question has been posed in the context of viral marketing \cite{domingos2001mining} but has focused on diffusion under latent, non-explicit preferences and, thus, non-explicit costs \cite{kempe2003maximizing}. 
Incentive schemes that help early adopters overcome initial barriers can initiate a cascade of behavioral change \cite{otto2020social}. Interventions can directly incentivize behavior change, but they can also have indirect effects when behaviors are socially interdependent. In such situations, social processes, such as adherence to the social norms in one's networks, can amplify the direct effects of circumscribed interventions, thereby reducing the total cost of behavior change \cite{gavrilets2020dynamics, constantino2022scaling}. Empirical work has shown that small, strategically designed interventions can lead to social tipping points that trigger large-scale behavioral transformations \cite{andreoni2021predicting}. 

However, identifying “who to target” to achieve these tipping points is not straightforward \cite{hu2018strategies}. Real-world collective action problems often occur in complex social networks of heterogeneous individuals, where locally held social norms or deeply lodged technologies can reinforce the status quo \cite{constantino2022scaling, efferson2020promise}. Targeted individuals may face significant countervailing influences from their immediate social network \cite{centola2021influencers,van2024transformation}. Additionally, their preferences can increase their vulnerability to the first-mover dilemma \cite{andreoni2021predicting}. As a result, unsustained behavioral change within the targeted group can potentially hinder broader spillover effects. Further, heterogeneity influences whether social norms remain stable or shift \cite{gavrilets2020dynamics, efferson2020promise}, and network segregation and clustering of preferences hinder the diffusion of desirable behaviors \cite{vasconcelos2021segregation, ehret2022group,aral2013engineering}. Policymakers attempting to "seed" broad behavior change through circumscribed interventions additionally face several binding constraints---including budgetary, the speed of the behavioral transition, and the distributional consequences of intervening. They must thus consider how the size and target of interventions in contexts with heterogeneous preferences and different social network structures impact their speed, efficiency, and equity.

Prior work has shown that the promise of a circumscribed intervention to create endogenous social change resulting in social tipping depends critically on how preferences are distributed in a population and the social network structure \cite{efferson2020promise,nejad2015success,centola2018behavior,centola2021influencers,van2024transformation,han2014balanced}. This work shows that how an intervention is targeted---e.g. whether it focuses on those who are more resistant or amenable to change, or whether it is randomly distributed has implications for the extent of its direct and indirect effects on behavior. For interventions focusing on the underlying social network, the literature shows targeting peripheral or more central nodes or a balanced mix can be effective in propagating change. Building on threshold models of collective behavior, which explain how individual decisions accumulate into large-scale shifts \cite{granovetter1978threshold}, recent research suggests that behavioral cascades often emerge from complex contagion processes \cite{centola2007complex, centola2018behavior}. In the context of behavior change or technology adoption, complex contagion describes a process where multiple sources of social proof or "exposure" are necessary before an individual alters their own behavior. The spread of important sustainable behaviors, such as the adoption of rooftop solar, have been described as complex contagion processes \cite{wolske2020peer}, in part because they entail the adoption of new, unknown, and potentially risky technologies. Such processes depend strongly on network structures since behaviors only gain momentum when nodes receive multiple confirming signals \cite{centola2021influencers, van2024transformation}.

These studies tend to focus on where or how to optimally seed behavior change to initiate the greatest overall change in the social system, while ignoring constraints faced by practitioners considering interventions, such as costs and time constraints or fairness considerations. In contrast, prior work that does consider how to maximize influence with seeding strategies on networks under budget constraints, tends to include simplifying assumptions, such as homogeneous preferences \cite{ bakshy2011everyone,han2014balanced,de2020efficient}. Few, if any, studies consider how to optimally seed widespread behavior change using circumscribed interventions while considering both heterogeneous preferences and different network structures and the real-world constraints faced by policymakers. This presents a critical gap: While research suggests that transitions can be catalyzed through strategically targeted interventions, how to maximize the efficiency, equity, speed or durability of these interventions under different assumptions about preference heterogeneity and network structure has received limited consideration. 

The present study addresses this gap by employing a game-theoretic and computational modeling framework to identify incentive strategies that promote sustainable technology adoption in a cost-effective and context-sensitive manner. Specifically, we use this modeling framework to ask: given information about the network structure and distribution and clustering of preferences, which intervention strategy is most cost-effective, timely, and fair? We consider different preference distributions, ranging from uni- to bimodal, and examine a variety of network topologies (e.g., heterogeneous Barabási-Albert (BA) and homogeneous Erd\H{o}s-Renyi (ER) networks, modular and non-modular). The cost of incentivizing an individual depends on their specific preference and their position in the network, and thus the targeting strategy determines the total cost of intervention. We implement different targeting heuristics relying on information about individual preferences and local network properties to identify low-cost strategies. The efficiency of these strategies is evaluated in terms of total cost, time to reach a target adoption level (90\%), and the inequality of incentive distribution as measured by the Gini coefficient. We show which strategies are most cost-effective and equal for a given context by systematically exploring how resistance to change and segregation levels influence outcomes. Our results offer valuable insights for policymakers seeking to leverage population information to customize incentive schemes, thus, bridging the gap between theoretical insights and practical strategies for guiding societies toward more sustainable futures \cite{bak2021stewardship}. 
    % Create a "code name" for each figure. In the maintext, when you want that figure to appear in the manuscript call that function (e.g. \figCodeName).

%%%% Create and format Figure with subpanels: %%%% 
%\generateFigSubpanels[optionalRotationDegrees]{figCodeName}{figureNamePath}{size:propOfTextwidth}
%     {main caption}
%    {{
%         {subcaption1},
%         {subcaption2},
%         {subcaption3}
%     }}

%Then in the reference your figures with \cref{fig:figCodeName} and Reference subpanels with \cref{fig:figCodeName:A}

%%%% Function to create and format figures without subpanels %%%%

\generateFig{figa}{fig1.jpg}{1}
     {Strategies need to account for the distribution of preferences}
     {The most cost-effective strategies to achieve 90\% adoption are plotted for different preference distributions modeled using a transformed Beta distribution, \(\text{Beta}(\alpha, \alpha)\), symmetrically centered around the average preference strength. The parameter shape parameters are the same, and \(\alpha\) controls homogeneity. In A, the preference distributions are plotted corresponding to 5 points in the parameter space depicted using stars in B with matching colors. We test strategies based on information about preference distribution (B), networks (C), and the random targeting strategy. We then compare all strategies in D. Among strategies based on preferences (B) the amenable strategy is cost-effective across all preference distributions. In C, we see that in case of high resistance to change, it is better to target the highly connected nodes while targeting peripheral nodes when the population is more amenable to change. In D, we see that as preference distribution gets more heterogeneous, the amenable strategy outperforms network-based strategies over a greater range of average preferences. The population size is 1000 and is connected via a Barabási-Albert network (min $k$=10) with social influence parameter $\omega  =0.5$. }

\generateFig{figb}{fig2.jpg}{1}
     {Optimal strategies in segregated and clustered populations}
     {The most cost-effective strategies to achieve 90 \% adoption are plotted for varying average preferences for change and levels of homophily in the network. In A, networks corresponding to corner cases and the center of the parameter space are depicted. We test strategies based on information about preference distribution (B), networks (C), and the random targeting strategy. We then compare all strategies in D. In B, we see that while amenable strategy works for low homophily networks, for more segregated networks random strategy works better in case of high resistance while targeting resistant individuals works when the population is more amenable on average. In C, we see that homophily doesn't change the ordering of network-informed strategies. In D, we see that as the networks get more segregated, the network-informed strategies outperform preference-informed strategies over a greater range of average preferences. The population size is 1000 and is connected via a homophilous Barabási-Albert network (min $k=10$) with social influence parameter $\omega = 0.5$ and the preference distribution is given by a transformed \(\text{Beta}(\alpha, \alpha)\), with $\alpha=2$. }

\generateFig{figc}{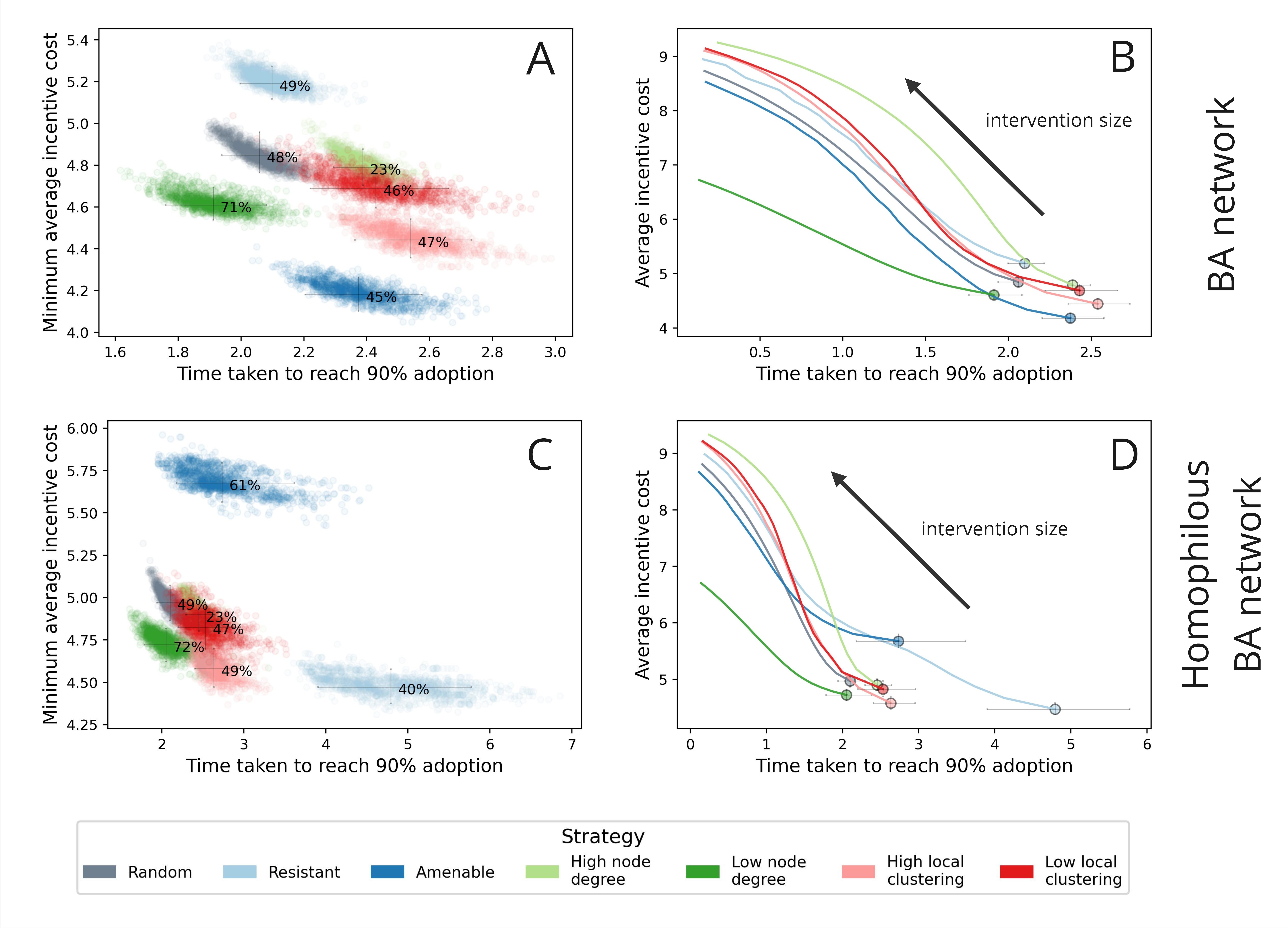}{1}
     {Time taken for 90\% adoption}
     {The time taken to reach 90\% adoption is plotted against the cost of implementing different heuristics for Barabási-Albert networks (A, B) and Barabási-Albert networks with homophily = 0.45 (C, D). In A, C the cost and time taken by minimum intervention using different strategies is plotted. Time is measured in generations ($N$ time steps). Each point represents the average value for a given configuration of network and preference distribution over 50 replicates. We consider 2500 configurations for each heuristic (50 networks and 50 preference distributions). The 5-95 percentile bands are shown along both axes. The average minimum interventions sizes which are indicated as percentages of the population. In B(D), as intervention sizes are increased beyond the minimum levels observed in A(C), the cost vs time plot shows an inflection point for all the tested heuristics. In the homophilous networks, all strategies take more time to achieve 90\% adoption. Further, the resistant strategy becomes cost-effective while taking significantly more time. The population size is 1000, with a  preference distribution given by a transformed \(\text{Beta}(\alpha, \alpha)\), with $\alpha=2$ centered around average preference = 1. }

\generateFig{figd}{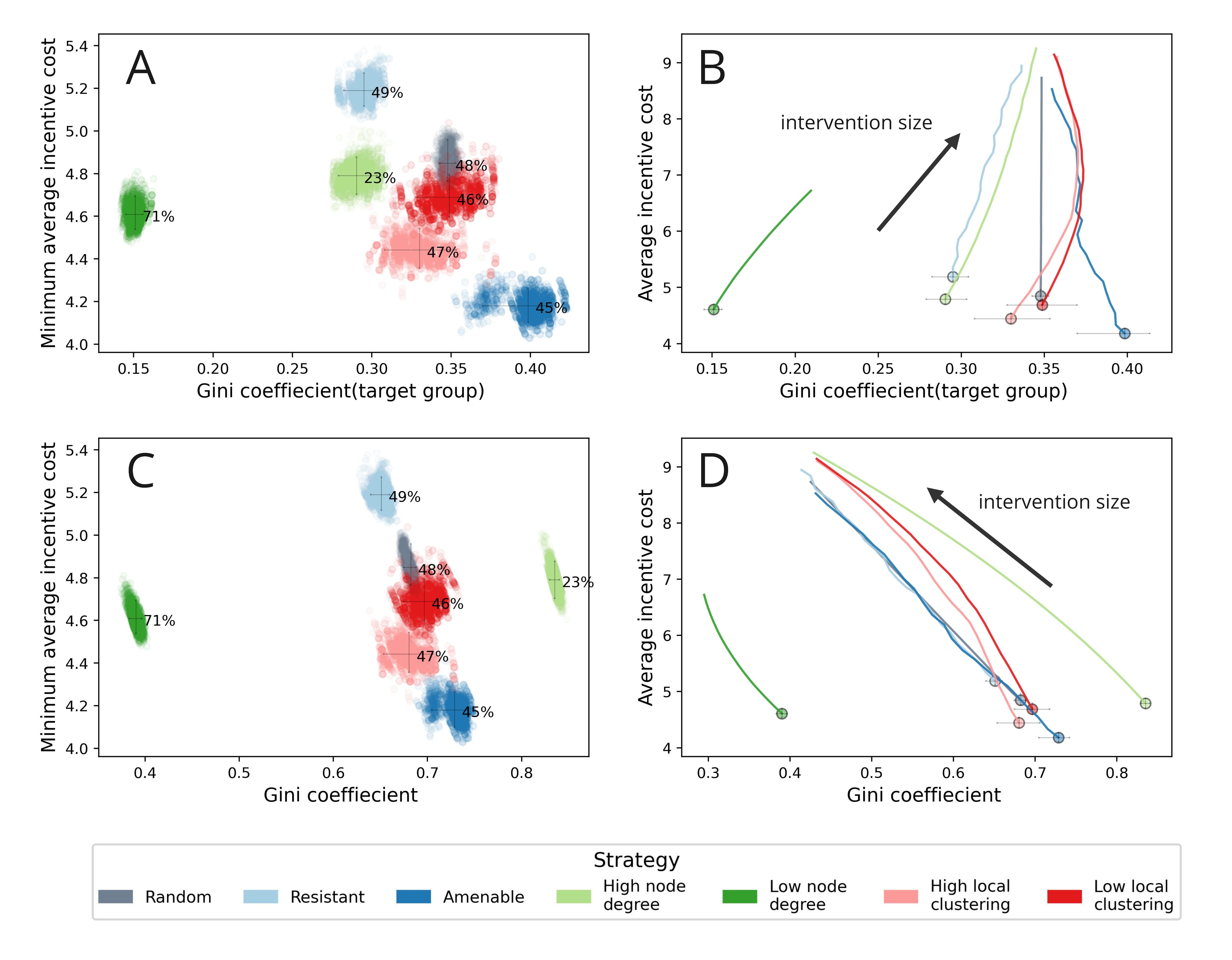}{1}
     {Gini coefficient of incentives}
     {The Gini coefficient of the incentives given is plotted for the target group (A, B) and the whole population (C, D).In A and C, the Gini coefficient is plotted against the cost of the minimum intervention sizes. As intervention sizes are increased beyond the minimum levels observed in \cref{fig:figc}. In B we see that the Gini coefficient within the target group increases with greater intervention size for network-informed strategies, and converges to the random strategy which remains unaffected by increasing intervention size being representative of the population heterogeneity. Thus smaller target groups are more relatively more homogeneous. In D, we see the intuitive result that greater intervention size decreases the Gini coefficient for the population( target and non-target group). The population size is 1000 connected by a Barabási-Albert network, with a  preference distribution given by a transformed \(\text{Beta}(\alpha, \alpha)\), with $\alpha=2$ centered around average preference = 1. }  

%\begin{figure}[!ht]
%\centering
%\generateSubFig{figa}{Fig_2_a.jpeg}{All strategies}{0.3}
%\generateSubFig{figb}{Fig_2_b.jpeg}%{Preference distribution}{0.3}
%\generateSubFig{figc}{Fig_2_c.jpeg}{Network properties}{0.3}
%\caption{Comparison of minimum interventions across different subsets of strategies.}
%\label{fig:strategy_comparison}
%\end{figure}

\generateFig{tablea}{table.PNG}{0.8}
     {Cost-optimal strategies based on resistance to change, network type, and segregation.}
     {* For uniform (bimodal) distribution of preferences- targeting high-degree (amenable) individuals is the best strategy. ** Beyond a certain level of segregation, the communities are separated enough for the resistant strategy to achieve spillover in the amenable community }
 %Then in the main text reference your figures with \cref{fig:fig1a}

% You can also use \generateSidwaysFigSubpanels or \generateSidewaysFig to rotate the caption with the figure.

%%%% Example figure with codename: figureSuggestions %%%%

\generateFigSubpanels{figureSuggestions}{figureSuggestions.pdf}{1}
    {Figure title which should not end in a period} % Main caption
    {{
        {Succinct caption for Panel A. Note that figure captions will not split across pages.}, % Subcaption 1
        {Caption for panel B. Add additional subpanel captions as needed depending on the figure.} % Subcaption 2
    }}
\section{RESULTS}
%\resultsTips % Delete when no longer needed

We consider a population of individuals, each characterized by intrinsic preferences towards a specific technology and local benefits to conform. Preferences and local conformity create, at a given time, a distribution of incentives that could be provided to each agent individually, which would make them take a set option. The difference between the minimal incentive to change and the option cost is the individual's willingness to pay. Thus, as adoption starts, willingness to pay will change as local pressures to conform change, even as intrinsic preferences remain fixed. This endogenous change in willingness to pay may lead to self-sustained adoption of a technology. For simplicity, we take these intrinsic preferences and the network of influences as fixed and, thus, as the defining characteristics of the population.

We focus on optimizing the cost of seeding a subset of early adopters. We test different targeting heuristics based on personal preferences and local network properties, namely, degree centrality and local clustering. We also test a random targeting strategy. In supplementary material (Fig S2), we also test strategies combining preferences and network structures \cite{bakshy2011everyone}. For each population configuration, we identify the minimum intervention size for all strategies required to achieve 90\% of technology adoption reliably. The strategy which, on average, costs the least is designated as the cost-effective strategy for that set of parameters. This analysis is done with an agent-based model and a mean-field model.

\figa

In the first scenario, we consider a population of individuals who influence each other in a Barabási-Albert heterogeneous network where a few nodes have a lot of connections, and most have few \cite{barabasi1999emergence} and examine the role of preference distribution in determining the cost-effectiveness of strategies. We vary the homogeneity of the distribution and the average relative preference for the new option. The preference distributions are modeled using a transformed Beta distribution,  \(\text{Beta}(\alpha, \alpha)\), symmetrically centered around the average preference strength. The parameter \(\alpha\) controls homogeneity, and by tuning it we get distributions ranging from unimodal to bimodal.

First, we consider a situation in which the intervenor only has information regarding the preference distribution to determine the cost-effective strategy in \cref{fig:figa}.  Among the preference-informed strategies and the random strategy, targeting amenable individuals---individuals with a high preference for change---emerges as the cost-effective strategy across all tested preference distributions \cref{fig:figa}B. In the supplementary, we show this analytically with a Markov chain analysis.

When the intervenor only has information about the network topology, they can apply network-informed strategies (\cref{fig:figa}C). In that case, targeting agents with high(low) degree-centrality is most effective when the population is highly resistant(amenable) to change---i.e., has a strong preference for the current (new) option. For intermediate resistance targeting nodes of high local clustering is the best strategy, providing an optimal combination of low-degree and high-degree nodes (Fig S4). Thus, choosing the appropriate direction of technology propagation, i.e., from the center to the periphery, vice-versa, or emanating from in between the two, can help minimize cost depending on the resistance in the population. 
\figb
Suppose the intervenor has both types of information, on comparing all strategies together(\cref{fig:figa}D). In that case, the amenable strategy emerges as the best strategy over a big part of the tested parameter space, especially when there are no strong preferences for either and there is high heterogeneity in preferences. 

These results hold as we vary the heterogeneity of networks except in more homogeneous networks, the amenable strategy outperforms the network-based strategy for all average preference strengths (Fig S5). Further, simulations on empirical networks \cite{jeon2015us} yield similar results(Fig S6).

 These findings align with the idea that the "influentials hypothesis" has a limited range of validity \cite{watts2007influentials}. To further explore this, we test a variation of the model where agents who switch to the new choice cannot revert to the previous one (Fig S7). In this case, targeting high-degree individuals becomes cost-effective over a larger parameter space. This suggests that while influential individuals have greater reach, they are also more susceptible to social push-back from their numerous connections, which can lead them to backslide when early adoption is possible . Conversely, peripheral individuals face less social resistance but have limited reach, making this strategy effective only when the population strongly prefers change. The amenable strategy remains robust because it targets a mix of influential and peripheral individuals  \cite{han2014balanced} who, due to their stronger preference for change, are less likely to revert to the status quo. In a similar vein, with no backsliding, the strategy of targeting resistant individuals also gains ground. Thus a strategy that can potentially generate large spillovers may also increase the risk of reversal within the target group due to greater exposure to opposing social influences \cite{centola2021influencers,van2024transformation}. This underscores the trade-off between maximizing reach and ensuring sustained adoption.

Further evidence of this trade-off emerges when we vary the degree assortativity in Barabási-Albert networks (Fig S8). As the network becomes more disassortative—meaning high-degree nodes are more likely to connect with low-degree nodes—targeting hubs becomes less effective. This may be due to the relatively small target group size, which, in the absence of strong in-group reinforcement, becomes more vulnerable to backsliding. On the other hand, targeting peripheral individuals gains effectiveness over a wider parameter space, as their reach improves while maintaining enough in-group links to sustain the change. In highly assortative networks, where high-degree nodes tend to connect primarily with other high-degree nodes, both degree-based strategies become less efficient, highlighting the complex interplay between network structure and intervention success.  

\figc
In the second scenario, we study the cost-effectiveness of strategies when the population is segregated based on preferences \cref{fig:figb}. We use homophilous Barabási-Albert graphs to generate networks with increasing levels of segregation between the two communities. This can be viewed as a proxy for populations of varying issue-based and affective polarization scenarios \cite{vasconcelos2021segregation}. 

Once again, we begin with an intervenor that only has information about the preferences of individuals. (\cref{fig:figb}B) As the level of homophily increases and the bridge between the communities becomes weaker, interventions in the amenable community cannot produce spillover in the resistant population. Targeting resistant individuals becomes the more effective strategy when the population, on average, favors change. However, in the face of greater resistance to change, the bridge between communities becomes dysfunctional, and targeting randomly outperforms the preference-informed strategies. We see that as the preferences become sorted in the network, for high average preference for change, the cost of implementing preference-based strategies decreases (Fig S9). This can be because the clustering of early adopters helps prevent them from backsliding by providing reinforcement.

Segregation and clustering of preferences make preference-based strategies less effective compared to the network-based strategies which target both communities, thus, compensating for the weak bridge between the communities. For highly resistant populations, we recover the classical result that targeting high-degree nodes is a good strategy \cite{chen2009efficient}.
\figd

Beyond minimal costs, each strategy has different implications. We now look into other essential considerations of interventions-- the time taken to achieve 90 \% adoption, the heterogeneity of incentives distributed, and the number of people incentivized for the different targeting strategies. In \cref{fig:figc} we show the time-effectiveness of strategies for population connected by a BA network with a particular distribution of preferences (average preference = 1 and $\alpha=2$). Targeting high degree agents generates the biggest spillovers while the intervention size required is largest while targeting peripheral agents. Further, we increased intervention sizes for the different strategies to estimate the trade-off between the extra cost incurred and the reduction in time taken to achieve 90\% adoption (\cref{fig:figc}B). The cost vs time plot follows a convex curve for the tested strategies, showing that beyond certain intervention sizes the costs pile up without substantial gains in terms of time. This happens at similar rates for different strategies. 

We compare these results with populations with the same preference distribution but segregated according to preferences (\cref{fig:figc}C, D). All strategies take more time to achieve 90\% adoption. Further, targeting the resistant individuals while being cost-effective takes much longer on average than the other strategies and it requires targeting individuals in the amenable subpopulation to make substantial gains in time-effectiveness.

Next, we look at the equity of incentives in the target group and in the entire population which also considers those that are not incentivized by calculating the Gini coefficient in the incentives given out \cref{fig:figd}. This analysis is done for the same population setup used in \cref{fig:figc} A, B. Targeting peripheral is the most equitable distribution of funds, because of the homogeneity of degree within this subgroup, the size of intervention required and also because of the relative small incentive required per individual. On the other hand, hubs face the most social pressure and, need the biggest incentives. Targeting hubs, thus concentrates the funds in a much smaller section of the population, making this strategy less equitable (\cref{fig:figd} C). The rest of the strategies, on the other hand, require a much bigger seeding base which results in a more equitable distribution of funds.  

The Gini coefficient within the target group increases with greater intervention size for network-informed strategies and converges to the random strategy which remains unaffected by increasing intervention size being representative of the population's heterogeneity in willingness to pay (\cref{fig:figd} B). Further, as intuition suggests, interventions targeting bigger groups are more equitable when considering the entire population (\cref{fig:figd} D).

We also test a version of the model in which the influence of a neighbor is dependent on the degree of the focal node--- individuals with fewer connections (low node degree) place higher importance on each neighbor's behavior and individuals with a higher node degree assign less weight to each neighbor (Fig S10). This model is representative of cases when the social influence is psychological. In such cases, targeting central individuals maximizes spillover effects without incurring additional costs.

    \section{DISCUSSION}
%\discussionTips % Delete when no longer needed
%In this paper, we explore the cost optimization of seeding early adopters in heterogeneous, networked populations by leveraging information about preference distributions and the social network structure. We contextualize the costs of interventions using a game-theoretic model, which enables us to identify targeting heuristics aimed at minimizing costs. This approach also informs the fairness of incentive distribution schemes and the timeliness of intended social transformations.

Our model is designed for cases where individuals directly benefit from coordination with one's neighbors and can easily revert their behavior after intervention. All individuals are equally influenced by their neighbors and this can be representative of cases where social influence is economic or psychological in nature. Our findings suggest that understanding the direction in which the behavioral change tends to be propagated on a network after intervention becomes crucial for cost minimization. Cost optimality is achieved by striking a balance between the self-reinforcement within the target group and the indirect influence that can be achieved, minimizing the risk of backsliding. The heterogeneity and clustering of preferences and of the network, thus, determine the cost-optimal targeting strategy. Further, as the average strength in preferences varies, and consequently the required intervention sizes, the efficiency of a target group also varies based on its composition.

%Given that different individuals require different levels of incentive and that they can revert to the old behavior in the absence of reinforcement from peers, the question of who to target becomes non-trivial. For cost-effective and reliable wide-spread adoption, the direction of the social contagion, also plays an important role. Specifically, initiating the contagion in the center of the social networks to propagate outwards is shown to be more cost-effective in the case of high resistance to change in the population. On the other hand, initiating change in the network's periphery to propagate towards the center works better when the population is more amenable to change. Further in the case of high structural segregation, initiating change in more resistant individuals is more effective than targeting amenable individuals. 

Several simplifying assumptions underpin our model. First, we assume timescales to be short enough to neglect dynamic changes in preferences or networks. Second, we posit that the policymaker has access to some detailed micro-level information—an assumption that, while useful for modeling, is often unfeasible in large-scale settings and raises potential ethical concerns. We distinguish cases of information about the network and information about individuals' preferences. We do not consider push-back effects from the interventions or differential responses to interventions, effects observed in real-world social norm interventions \cite{efferson2020promise,efferson2024norm}.

Further, incentives tailored to suit individual attitudes can be exploited by individuals who may indicate a stronger aversion to change to get bigger incentives. In such a case, flat incentives \cite{ehret2022group} would be essential and the magnitude of incentives would be dependent on the target group . 
  
While our model assumes a linear relationship between utility and the frequency of choices in the neighborhood, real-world dynamics can be nonlinear. For instance, in the context of electric vehicle adoption, a neighborhood might only install charging stations after a critical mass of residents own electric cars. Further, we assume that all individuals are equally affected by social pressure. This assumption can be relaxed for more realistic scenarios to account for heterogeneity, i.e., having susceptible, insusceptible/non-conforming, or even anti-conforming individuals \cite{hu2018strategies,mittal2024anti,pinheiro2025heterogeneous}. This presents an additional layer of complexity that affects the efficacy of targeting strategies while also presenting a different set of potential strategies accounting for conforming tendencies.

Literature based on the threshold model suggests targeting resistant individuals to maximize spillover effects and random targeting for more robust outcomes \cite{efferson2020promise}. Our results, however, suggest that, when considering total costs and the possibility of backsliding, targeting amenable individuals can be a better strategy. This apparent contradiction arises because the study in mention assumes that the cost of changing an individual's behavior is the same for all individuals, while we specifically do away with this simplification. We plot how different strategies map on the effective threshold distribution for a direct comparison (Fig. S11). This underlines the need to contextualize thresholds to particular problems, behaviors, or technologies to gain more useful insights, particularly for designing real-world interventions \cite{guilbeault2018complex}.

With cost considerations, we identify population setups where different network-based strategies, like the billboard (targeting central individuals) \cite{granovetter1973strength}, the handbill (targeting peripheral individuals) \cite{van2024transformation} and the balanced targeting \cite {han2014balanced}, are optimal \cite{centola2018behavior,centola2021influencers,van2024transformation}. 

% This problem has been studied in network science, computer science, and product marketing research as the Influence Maximization problem. However,  most studies assume homogeneity or consider thresholds independent of individuals' social context. With our model, we aim to contextualize an individual's thresholds in terms of personal preference, conforming pressures, and the position she occupies in the social network. This also provides a natural setting for a more heterogeneous distribution of thresholds. Further, our model accounts for competition between the two options --- it assumes symmetry between options rather than a ratcheted system where non-adopters of the new technology do not actively influence the decision-making process.  
In our work, our goal has been to understand the targeting potential of individuals and relate it to simple characterizations of these individuals. We have, thus, focused on strategies driven by a single criterion, but more sophisticated targeting using information about the preference distribution and the network can yield greater optimization. We try out a strategy that targets individuals with a high ratio of social influence to the cost of incentive. This strategy outperforms other strategies for a considerable part of the parameter space, but becomes inefficient in the case of network segregation. Similarly, other optimal combinations can potentially be worked out for different population setups. Further, we incentivize all targeted individuals simultaneously in the beginning. This setting is often studied in the literature of complex networks in the context of seeding strategies, but it has focused both on irreversible processes (such as information propagation) and homogeneous nodes (apart from degree). A more nuanced approach accounting for an optimal dynamic targeting \cite{wang2023optimization} can offer even more cost-effective solutions.

While our study provides theoretical insights, these can be empirically tested in experimental setups. Future research could explore feedback loops between behavior and environmental changes \cite{tilman2020evolutionary}, where shifts in the environment influence individual preferences in turn.

% However, accurately mapping preferences and network positions can be difficult, and policymakers may have to design targeting schemes with the information available to them. Our model shows that targeting strategies making use of maximum information perform better. However, in the lack of complete information, policymakers can still make informed decisions that can yield better outcomes than random targeting. Further, this can also help policymakers identify scenarios where information about the network or the preferences would help them to make effective interventions.
% In smaller populations or in organizations, identifying the underlying influence network can be easier.

    \section{METHODS}
%\methodsTips % Delete when no longer needed
 \subsection{Model}
We develop an Agent-Based Model (ABM) of collective behavior in which a population of $N$ individuals adopt one of two different products, $A$ or $B$. At each time step, an individual is asynchronously selected to consider changing the product they currently use. The decision process is
based on a game-theoretic framework of collective decision-making, which captures (reversible) complex contagion behavior from microscopic principles \cite{mittal2024anti,gavrilets2020dynamics,yang2022sociocultural}. An agent $i$ faces two options and makes decisions that maximize their utility. The utility of each choice is characterized by net intrinsic individual preferences $o_i^A$ and $o_i^B$ for products $A$ and $B$, and by a term derived from the products adopted in their social network \cite{gavrilets2020dynamics}. The decision weighs both, i.e., $\Delta o_i=o_i^A-o_i^B,$ along with the net influence exerted through the immediate social network, $\omega\,(\#_i^A-\#_i^B)$, where $\#_i^A$ ($\#_i^B$) denotes the number of neighbors of $i$ choosing A (B). The net social influence can be described in terms of the benefits derived from coordination on a choice or in terms of the penalty received in the case of miscoordination. The two cases are equivalent. The utility of option $X\in\{A,B\}$ is $U_i^X=o_i^X+\omega \, \#_i^X$. Thus, the marginal utility of choice A, $\Delta U_i\equiv U_i^A-U_i^B$, is

\begin{equation}
    \Delta U_i = \Delta o_i + \omega\,( {\# _i}^A -{\# _i}^B).
\label{delta_U}
\end{equation}

When $\Delta U_i$ is positive (negative) agent $i$ chooses A (B). Thus, each agent is characterized by a threshold number of As in their neighborhood above which they will act A and below which they will act B. That threshold is simply $T_i=k_i/2-\Delta o_i/(2\omega)$. Agents are also characterized by a benefit to action A above which they would adopt action A, given by $b_i=\omega(k_i- 2{\# _i}^A ) -\Delta o_i$. Further, agent's choices are reversible, i.e., they can change their choice if they do not experience enough reinforcement in their neighborhood.

\subsection{Analytical methods}
We complement the ABM with analysis using Markov chains \cite{traulsen2005coevolutionary} and calculate the transition probability of the macro-state using a mean-field approach based on the probabilities of switching choices in all the different sub-populations for the targeted and non-targeted group.

 \subsection{Intervention implementation}
 The targeting strategies tested were based on individual preferences and network properties, namely degree centrality and local clustering centrality. We chose node degree as the centrality measure as other measures like between-ness centrality and eigenvector centrality are highly correlated with node degree for the networks we used. 

 The agents are first sorted according to the property on which the strategy is based and then the individuals from the top (or bottom) of the sorted list are selected based on the intervention size.  The utility of the new technology perceived by targeted individuals is determined and accordingly, they are given personalized incentives that are just enough to tip them over the point of indifference. This is done at the start of the simulation and then the collective decision-making is allowed to take over. This helps us identify the minimum interventions required to achieve social transition reliably, i.e., when the intervention leads to 90 \% adoption or more for at least 75 \% of all the population configurations. The analysis is robust to the target fraction. (Fig S3)

 For a given set of seeded individuals, X, the cost of the intervention is given by $\sum_{i\in X} b_i$ with $\#_i^A=0$.

 \subsection{Networks}
  \subsubsection{Homophilous networks}
 To generate homophilous networks, we sort the population according to preferences and split it into two communities. Then we modify the preferential attachment algorithm to include the node's preference to form an edge with another node from the same community. The homophily parameter tunes this preference of attaching within the group, for homophily = 0, there is an equal chance of connecting with nodes in both communities, while for homophily = 1, a node only connects within the same community.
  \subsubsection{Networks with varying heterogeneity}
To generate networks, ranging from homogeneous( Poisson degree distribution) to heterogeneous (scale-free degree distribution) we calculate the weighted average of the two probability distributions of degree, tune the weights and then use a configuration model to generate networks with the corresponding degree distribution. This helps us tune the heterogeneity of the networks using one weight parameter.
  \subsubsection{Networks with varying degree-assortativity}
Assortative networks are generated using a reshuffling algorithm \cite{xulvi2004reshuffling}. 
  \subsubsection{Empirical networks}
We use 74 networks from the Add Health Dataset \cite{jeon2015us,guilbeault2021topological}.

 \subsection{Data and Code Availability}
The code used for the simulations can be accessed at \url{https://github.com/d-mittal/cost_optimal_incentives}. The static version of the code will be made available as a Zenodo repository.

 \subsection{Acknowledgements}
D.M. and V.V.V. acknowledge funding from the Computational Science Lab and ENLENS under the project "The Cost of Large-Scale Transitions: Introducing Effective Targeted Incentives."

%\subsection{Author Contributions}
%Use the CRediT Taxonomy to indicate author contributions. See \url{https://www.cell.com/pb/assets/raw/shared/guidelines/CRediT-taxonomy-1430242873507.pdf}.

 \subsection{Competing Interests}
The authors declare no competing interests.
    \clearpage
    \begin{singlespace}
        \printbibliography

@article{tilman2020evolutionary,
  title={Evolutionary games with environmental feedbacks},
  author={Tilman, Andrew R and Plotkin, Joshua B and Ak{\c{c}}ay, Erol},
  journal={Nature communications},
  volume={11},
  number={1},
  pages={915},
  year={2020},
  publisher={Nature Publishing Group UK London}
}

@article{wang2023optimization,
  title={Optimization of institutional incentives for cooperation in structured populations},
  author={Wang, Shengxian and Chen, Xiaojie and Xiao, Zhilong and Szolnoki, Attila and Vasconcelos, V{\'\i}tor V},
  journal={Journal of the Royal Society Interface},
  volume={20},
  number={199},
  pages={20220653},
  year={2023},
  publisher={The Royal Society}
}

@article{pinheiro2025heterogeneous,
  title={Heterogeneous Update Processes Shape Information Cascades in Social Networks},
  author={Pinheiro, Fl{\'a}vio L and Vasconcelos, V{\'\i}tor V},
  journal={arXiv preprint arXiv:2501.08498},
  year={2025}
}

@article{barabasi1999emergence,
    title = {Emergence of Scaling in Random Networks},
author = {Barab\'{a}si, Albert-L\'{a}szl\'{o} and Albert, R\'{e}ka},
    journal = {Science},
 volume={286},
  pages={509-512},
    year = {1999} 
}

@article{traulsen2005coevolutionary,
  title={Coevolutionary dynamics: from finite to infinite populations},
  author={Traulsen, Arne and Claussen, Jens Christian and Hauert, Christoph},
  journal={Physical review letters},
  volume={95},
  number={23},
  pages={238701},
  year={2005},
  publisher={APS}
}

@article{wolske2020peer,
  title={Peer influence on household energy behaviours},
  author={Wolske, Kimberly S and Gillingham, Kenneth T and Schultz, P Wesley},
  journal={Nature Energy},
  volume={5},
  number={3},
  pages={202--212},
  year={2020},
  publisher={Nature Publishing Group UK London}
}

@article{pizziol2024niches,
  title={From niches to norms: the promise of social tipping interventions to scale climate action},
  author={Pizziol, Veronica and Tavoni, Alessandro},
  journal={npj Climate Action},
  volume={3},
  number={1},
  pages={46},
  year={2024},
  publisher={Nature Publishing Group UK London}
}

@article{constantino2022scaling,
  title={Scaling up change: A critical review and practical guide to harnessing social norms for climate action},
  author={Constantino, Sara M and Sparkman, Gregg and Kraft-Todd, Gordon T and Bicchieri, Cristina and Centola, Damon and Shell-Duncan, Bettina and Vogt, Sonja and Weber, Elke U},
  journal={Psychological science in the public interest},
  volume={23},
  number={2},
  pages={50--97},
  year={2022},
  publisher={Sage Publications Sage CA: Los Angeles, CA}
}

@article{jeon2015us,
  title={US adolescents’ friendship networks and health risk behaviors: a systematic review of studies using social network analysis and Add Health data},
  author={Jeon, Kwon Chan and Goodson, Patricia},
  journal={PeerJ},
  volume={3},
  pages={e1052},
  year={2015},
  publisher={PeerJ Inc.}
}

@article{guilbeault2021topological,
  title={Topological measures for identifying and predicting the spread of complex contagions},
  author={Guilbeault, Douglas and Centola, Damon},
  journal={Nature communications},
  volume={12},
  number={1},
  pages={4430},
  year={2021},
  publisher={Nature Publishing Group UK London}
}

@article{xulvi2004reshuffling,
  title={Reshuffling scale-free networks: From random to assortative},
  author={Xulvi-Brunet, Ramon and Sokolov, Igor M},
  journal={Physical Review E—Statistical, Nonlinear, and Soft Matter Physics},
  volume={70},
  number={6},
  pages={066102},
  year={2004},
  publisher={APS}
}

@article{efferson2024norm,
  title={When norm change hurts},
  author={Efferson, Charles and Ehret, S{\"o}nke and von Fl{\"u}e, Lukas and Vogt, Sonja},
  journal={Philosophical Transactions of the Royal Society B},
  volume={379},
  number={1897},
  pages={20230039},
  year={2024},
  publisher={The Royal Society}
}

@article{guilbeault2018complex,
  title={Complex contagions: A decade in review},
  author={Guilbeault, Douglas and Becker, Joshua and Centola, Damon},
  journal={Complex spreading phenomena in social systems: Influence and contagion in real-world social networks},
  pages={3--25},
  year={2018},
  publisher={Springer}
}

@inproceedings{chen2009efficient,
  title={Efficient influence maximization in social networks},
  author={Chen, Wei and Wang, Yajun and Yang, Siyu},
  booktitle={Proceedings of the 15th ACM SIGKDD international conference on Knowledge discovery and data mining},
  pages={199--208},
  year={2009}
}

@article{watts2007influentials,
  title={Influentials, networks, and public opinion formation},
  author={Watts, Duncan J and Dodds, Peter Sheridan},
  journal={Journal of consumer research},
  volume={34},
  number={4},
  pages={441--458},
  year={2007},
  publisher={The University of Chicago Press}
}

@article{centola2007complex,
  title={Complex contagions and the weakness of long ties},
  author={Centola, Damon and Macy, Michael},
  journal={American journal of Sociology},
  volume={113},
  number={3},
  pages={702--734},
  year={2007},
  publisher={The University of Chicago Press}
}

@article{vasconcelos2015cooperation,
  title={Cooperation dynamics of polycentric climate governance},
  author={Vasconcelos, V{\'\i}tor V and Santos, Francisco C and Pacheco, Jorge M},
  journal={Mathematical Models and Methods in Applied Sciences},
  volume={25},
  number={13},
  pages={2503--2517},
  year={2015},
  publisher={World Scientific}
}

@article{santos2011risk,
  title={Risk of collective failure provides an escape from the tragedy of the commons},
  author={Santos, Francisco C and Pacheco, Jorge M},
  journal={Proceedings of the National Academy of Sciences},
  volume={108},
  number={26},
  pages={10421--10425},
  year={2011},
  publisher={National Acad Sciences}
}

@incollection{Creutzig2022,
  author       = {Creutzig, F and J Roy and P Devine-Wright and J Díaz-José and FW Geels and A Grubler and N Maïzi and E Masanet and Y Mulugetta and CD Onyige and PE Perkins and A Sanches-Pereira and EU Weber},
  title        = {Demand, Services, and Social Aspects of Mitigation},
  booktitle    = {Climate Change 2022: Mitigation of Climate Change. Contribution of Working Group III to the Sixth Assessment Report of the Intergovernmental Panel on Climate Change},
  editor       = {Shukla, PR and J Skea and R Slade and A Al Khourdajie and R van Diemen and D McCollum and M Pathak and S Some and P Vyas and R Fradera and M Belkacemi and A Hasija and G Lisboa and S Luz and J Malley},
  publisher    = {Cambridge University Press},
  address      = {Cambridge, UK \& New York, NY, USA},
  year         = {2022},
  doi          = {10.1017/9781009157926.007},
  note         = {\url{https://www.ipcc.ch/report/ar6/wg3/chapter/chapter-5/}}
}

@article{weber2006experience,
  title={Experience-based and description-based perceptions of long-term risk: Why global warming does not scare us (yet)},
  author={Weber, Elke U},
  journal={Climatic change},
  volume={77},
  number={1},
  pages={103--120},
  year={2006},
  publisher={Springer}
}

@article{yang2022sociocultural,
  title={Sociocultural determinants of global mask-wearing behavior},
  author={Yang, Luojun and Constantino, Sara M and Grenfell, Bryan T and Weber, Elke U and Levin, Simon A and Vasconcelos, V{\'\i}tor V},
  journal={Proceedings of the National Academy of Sciences},
  volume={119},
  number={41},
  pages={e2213525119},
  year={2022},
  publisher={National Acad Sciences}
}

@article{van2024transformation,
  title={Transformation starts at the periphery of networks where pushback is less},
  author={van de Leemput, Ingrid A and Bascompte, Jordi and Buddendorf, Willem Bastiaan and Dakos, Vasilis and Lever, J Jelle and Scheffer, Marten and van Nes, Egbert H},
  journal={Scientific Reports},
  volume={14},
  number={1},
  pages={11344},
  year={2024},
  publisher={Nature Publishing Group UK London}
}

@article{banerjee2013diffusion,
  title={The diffusion of microfinance},
  author={Banerjee, Abhijit and Chandrasekhar, Arun G and Duflo, Esther and Jackson, Matthew O},
  journal={Science},
  volume={341},
  number={6144},
  pages={1236498},
  year={2013},
  publisher={American Association for the Advancement of Science}
}

@inproceedings{han2014balanced,
  title={Balanced seed selection for budgeted influence maximization in social networks},
  author={Han, Shuo and Zhuang, Fuzhen and He, Qing and Shi, Zhongzhi},
  booktitle={Advances in Knowledge Discovery and Data Mining: 18th Pacific-Asia Conference, PAKDD 2014, Tainan, Taiwan, May 13-16, 2014. Proceedings, Part I 18},
  pages={65--77},
  year={2014},
  organization={Springer}
}

@inproceedings{bakshy2011everyone,
  title={Everyone's an influencer: quantifying influence on twitter},
  author={Bakshy, Eytan and Hofman, Jake M and Mason, Winter A and Watts, Duncan J},
  booktitle={Proceedings of the fourth ACM international conference on Web search and data mining},
  pages={65--74},
  year={2011}
}

@article{hu2018strategies,
  title={Strategies for new product diffusion: whom and how to target?},
  author={Hu, Hai-hua and Lin, Jun and Qian, Yanjun and Sun, Jian},
  journal={Journal of Business Research},
  volume={83},
  pages={111--119},
  year={2018},
  publisher={Elsevier}
}

@article{de2020efficient,
  title={Efficient network seeding under variable node cost and limited budget for social networks},
  author={De Souza, RC and Figueiredo, Daniel R and Rocha, AA de A and Ziviani, Artur},
  journal={Information Sciences},
  volume={514},
  pages={369--384},
  year={2020},
  publisher={Elsevier}
}

@article{aral2013engineering,
  title={Engineering social contagions: Optimal network seeding in the presence of homophily},
  author={Aral, Sinan and Muchnik, Lev and Sundararajan, Arun},
  journal={Network Science},
  volume={1},
  number={2},
  pages={125--153},
  year={2013},
  publisher={Cambridge University Press}
}

@article{nejad2015success,
  title={Success factors in product seeding: The role of homophily},
  author={Nejad, Mohammad G and Amini, Mehdi and Babakus, Emin},
  journal={Journal of Retailing},
  volume={91},
  number={1},
  pages={68--88},
  year={2015},
  publisher={Elsevier}
}

@article{santos2021biased,
  title={Biased perceptions explain collective action deadlocks and suggest new mechanisms to prompt cooperation},
  author={Santos, Fernando P and Levin, Simon A and Vasconcelos, V{\'\i}tor V},
  journal={Iscience},
  volume={24},
  number={4},
  year={2021},
  publisher={Elsevier}
}

@article{vasconcelos2021segregation,
  title={Segregation and clustering of preferences erode socially beneficial coordination},
  author={Vasconcelos, V{\'\i}tor V and Constantino, Sara M and Dannenberg, Astrid and Lumkowsky, Marcel and Weber, Elke and Levin, Simon},
  journal={Proceedings of the National Academy of Sciences},
  volume={118},
  number={50},
  pages={e2102153118},
  year={2021},
  publisher={National Acad Sciences}
}

@article{dietz2009household,
  title={Household actions can provide a behavioral wedge to rapidly reduce US carbon emissions},
  author={Dietz, Thomas and Gardner, Gerald T and Gilligan, Jonathan and Stern, Paul C and Vandenbergh, Michael P},
  journal={Proceedings of the National Academy of Sciences},
  volume={106},
  number={44},
  pages={18452--18456},
  year={2009},
  publisher={National Acad Sciences}
}

@article{nyborg2016social,
  title={Social norms as solutions},
  author={Nyborg, Karine and Anderies, John M and Dannenberg, Astrid and Lindahl, Therese and Schill, Caroline and Schl{\"u}ter, Maja and Adger, W Neil and Arrow, Kenneth J and Barrett, Scott and Carpenter, Stephen and others},
  journal={Science},
  volume={354},
  number={6308},
  pages={42--43},
  year={2016},
  publisher={American Association for the Advancement of Science}
}

@article{andreoni2021predicting,
  title={Predicting social tipping and norm change in controlled experiments},
  author={Andreoni, James and Nikiforakis, Nikos and Siegenthaler, Simon},
  journal={Proceedings of the National Academy of Sciences},
  volume={118},
  number={16},
  pages={e2014893118},
  year={2021},
  publisher={National Acad Sciences}
}

@article{otto2020social,
  title={Social tipping dynamics for stabilizing Earth’s climate by 2050},
  author={Otto, Ilona M and Donges, Jonathan F and Cremades, Roger and Bhowmik, Avit and Hewitt, Richard J and Lucht, Wolfgang and Rockstr{\"o}m, Johan and Allerberger, Franziska and McCaffrey, Mark and Doe, Sylvanus SP and others},
  journal={Proceedings of the National Academy of Sciences},
  volume={117},
  number={5},
  pages={2354--2365},
  year={2020},
  publisher={National Acad Sciences}
}

@article{bak2021stewardship,
  title={Stewardship of global collective behavior},
  author={Bak-Coleman, Joseph B and Alfano, Mark and Barfuss, Wolfram and Bergstrom, Carl T and Centeno, Miguel A and Couzin, Iain D and Donges, Jonathan F and Galesic, Mirta and Gersick, Andrew S and Jacquet, Jennifer and others},
  journal={Proceedings of the National Academy of Sciences},
  volume={118},
  number={27},
  pages={e2025764118},
  year={2021},
  publisher={National Acad Sciences}
}

@book{centola2018behavior,
  title={How behavior spreads: The science of complex contagions},
  author={Centola, Damon},
  volume={3},
  year={2018},
  publisher={Princeton University Press Princeton, NJ}
}

@article{granovetter1978threshold ,
  title={Threshold models of collective behavior},
  author={Granovetter, Mark},
  journal={American journal of sociology},
  volume={83},
  number={6},
  pages={1420--1443},
  year={1978},
  publisher={University of Chicago Press}
}

@article{efferson2020promise,
  title={The promise and the peril of using social influence to reverse harmful traditions},
  author={Efferson, Charles and Vogt, Sonja and Fehr, Ernst},
  journal={Nature human behaviour},
  volume={4},
  number={1},
  pages={55--68},
  year={2020},
  publisher={Nature Publishing Group UK London}
}

@article{gavrilets2020dynamics,
  title={The dynamics of injunctive social norms},
  author={Gavrilets, Sergey},
  journal={Evolutionary Human Sciences},
  volume={2},
  pages={e60},
  year={2020},
  publisher={Cambridge University Press}
}

@article{ehret2022group,
  title={Group identities can undermine social tipping after intervention},
  author={Ehret, S{\"o}nke and Constantino, Sara M and Weber, Elke U and Efferson, Charles and Vogt, Sonja},
  journal={Nature human behaviour},
  volume={6},
  number={12},
  pages={1669--1679},
  year={2022},
  publisher={Nature Publishing Group UK London}
}

@article{centola2021influencers,
  title={Influencers, backfire effects, and the power of the periphery},author={Centola, D},
  journal={Personal Networks: Classic Readings and New Directions in Egocentric Analysis},
  volume={51},
  pages={73},
  year={2021},
  publisher={Cambridge University Press Cambridge}
}

@article{mittal2024anti,
  title={Anti-conformists catalyze societal transitions and facilitate the expression of evolving preferences},
  author={Mittal, Dhruv and Constantino, Sara M and Vasconcelos, V{\'\i}tor V},
  journal={PNAS nexus},
  pages={pgae302},
  year={2024},
  publisher={Oxford University Press}
}

@article{granovetter1973strength,
  title={The strength of weak ties},
  author={Granovetter, Mark S},
  journal={American journal of sociology},
  volume={78},
  number={6},
  pages={1360--1380},
  year={1973},
  publisher={University of Chicago Press}
}
    \end{singlespace}
    \clearpage
    
    %\section*{Supplementary Information}
    \setcounter{page}{1} % Set page at 1
    \section*{Supplemental Text}

% %Some supplemental text that doesn't fit in the main text or provides extra detail can go here. We will use this space to outline some general tips.

%\subsubsection*{Analysis}

\subsection{Markov chain analysis of adoption under intervention}

We model the collective adoption dynamics in a well--mixed population by considering a Markov chain in which the state variable \(x\) denotes the overall fraction of individuals adopting the new behavior \cite{traulsen2005coevolutionary}. In our framework, the population is partitioned into two groups---targeted and non--targeted---via an intervention that selects a fraction \(\phi\) of the population according to a specified strategy (e.g., targeting individuals with extreme values of net preference \(\Delta o\) or degree \(k\)) or randomly. The dynamics of adoption in the two groups are then analyzed by considering the probability that a single update changes the state of the system  -- $x^T$ and $x^{NT}$ denoting the number of adopters in the target and non-target groups as a fraction of the entire population such that $x=x^T+x^{NT}$.

\subsubsection{Transition probabilities}

Let the distribution functions of preference and degree for the targeted group be $\rho_o^T,\rho_k^T$ and for the non-targeted group be $\rho_o^{NT},\rho_k^{NT}$. Similarly to the ABM, all agents have an equal likelihood of being selected at any given time, and we assume an independent distribution agent attributes preference ($\rho_o$) and degree ($\rho_k$) which corresponds to the case of random placement of preferences on the network. This gives us the transition probability for an agent changing behavior (i.e. changing \(x^T\) and \(x^{NT}\) by \(1/N\)). If in each time step we break the spatial correlations that build in the network (captured in the ABM), for the target group, the transition probabilities are given by
\[
T_{+}^T = (\phi-x^T) \int \!\!\! \int \rho_o^T \,\rho_k^T \, P_{o,k}\, do\, dk,\quad \text{and}\quad T_{-}^T = x^T \int \!\!\! \int \rho_o^T\, \rho_k^T \, \Bigl(1-P_{o,k}\Bigr)\, do\, dk \tag{S1}
\]
and similarly for the non--target group,
\[
T_{+}^{NT} = ((1-\phi)-x^{NT}) \int \!\!\! \int \rho_o^{NT}\, \rho_k^{NT} \, P_{o,k}\, do\, dk\, \quad \text{and}\quad T_{-}^{NT} = x^{NT} \int \!\!\! \int \rho_o^{NT} \,\rho_k^{NT} \, \Bigl(1-P_{o,k}\Bigr)\, do\, dk \tag{S2}
\]

where $P^{o,k}$ denotes the probability that an agent with net personal preference for change $\Delta o $ and degree $k$ adopts the new behavior

\[
P_{o,k} = \sum_{j=0}^{k} \binom{k}{j}\, p_{\text{link}}^j \, (1-p_{\text{link}})^{\,k-j}\, \Theta\Bigl(\Delta o + \omega(2j-k)\Bigr). \tag{S5}
\]

The parameter \(\omega\) represents the influence of observed behaviors or social conformity. The Heaviside function \(\Theta(\cdot)\) captures the threshold behavior. \(p_{\text{link}}\) denotes the probability that the focal agent has a link with an adopter and is given by 
\[
p_{\text{link}} = \frac{\, \langle k \rangle^T\, x^T + \, \langle k \rangle^{NT}\, x^{NT}}{\phi\, \langle k \rangle^T + (1-\phi)\, \langle k \rangle^{NT}}, \tag{S6}
\]

where $\langle k \rangle^T = \sum_{k} k\, \rho_k^T$ is the average degree in the target group and $\langle k \rangle^{NT} = \sum_{k} k\, \rho_k^{NT}$ is the average degree in the non-target group.

The resulting mean--field ODEs for the adoption dynamics in the two groups are
\[
\dot{x}^T =T_{+}^T - T_{-}^T= \phi \int \!\!\! \int  \rho_o^T\, \rho_k^T\, P_{o,k} do \, dk\, - x^T \tag{S3} \]

\[\dot{x}^{NT} = T_{+}^{NT} - T_{-}^{NT} = (1-\phi)\int \!\!\! \int  \rho_o^{NT}\, \rho_k^{NT}\, P_{o,k} do\,  dk - x^{NT}\tag{S4}
\]

\subsubsection{Amenable, Resistant and Random targeting}

For the targeting strategies which are agnostic to degree distributions: 
\[
\rho_k^T=\rho_k^{NT}=\rho_k \Rightarrow \langle k \rangle^T = \langle k \rangle^{NT}=\langle k \rangle \tag{S7}\label{sieq7}
\]
and hence $p_\text{link}=x^T+x^{NT}$.

For targeting amenable individuals:

\[\rho_o^T=\frac{\rho_o}{\phi}\,\mathds{1}_{o\geq o^*[\phi]}, \quad \rho_o^{NT}=\frac{\rho_o}{(1-\phi)}\,\mathds{1}_{o< o^*[\phi]} , \quad o^* : \int_{o^*}^{\infty}\rho_o do=\phi  \tag{S8}\]

the ODEs for $x^T$ and $x^{NT}$ become
\begin{align}
\dot{x}^T=\int\int \rho_o,\mathds{1}_{o\geq o^*[\phi]} \,\rho_k \, P_{o,k} [x^T+x^{NT}] do\,dk -x^T \tag{S9} \\
=\int_{0}^{\infty} \int_{o^*[\phi]}^{\infty} \rho_o \,\rho_k \, P_{o,k} [x^T+x^{NT}] do\,dk -x^T \nonumber
\end{align}

\begin{align}
\dot{x}^{NT}=\int\int \rho_o,\mathds{1}_{o< o^*[\phi]} \,\rho_k \, P_{o,k} [x^T+x^{NT}] do\,dk -x^{NT} \tag{S10}  \\ 
=\int_{0}^{\infty} \int_{-\infty}^{o^*[\phi]} \rho_o \,\rho_k \, P_{o,k} [x^T+x^{NT}] do\,dk -x^{NT}\nonumber
\end{align}

\begin{align}
    \dot{x}=\dot{x}^T+\dot{x}^{NT}=\int\int \rho_o \,\rho_k \, P_{o,k} [x] do\,dk -x 
\tag{S11}
\label{sieq11}
\end{align}

thus $\phi_{min}=x^*$ where $x^*$ is the largest unstable fixed point (of Eq. \ref{sieq11}) smaller than 1, which ensures complete adoption. Similarly for other strategies like targeting resistant individuals and random targeting which satisfy Eq. \ref{sieq7} it can shown that $\phi_{min}$ is determined by the unstable fixed point of Eq. \ref{sieq11}. Further, it is clear that among these strategies the amenable strategy is the most cost-effective as they share the same $\phi_{min}$ and amenable individuals cost less to incentivize by design.

By numerically integrating these ODEs, we theoretically analyze the trajectories of adoption under different intervention strategies and for various intervention sizes \(\phi\). This mean--field framework, supported by a Markov chain analysis in the large--population limit, provides insight into the critical thresholds and dynamical transitions observed in the system. 

In \cref{fig:theory}, we plot the trajectories of adoption in both the target and non-target groups for different intervention sizes and identify the minimum intervention size for targeting strategies based on preference distribution, degree distribution and a random strategy. The minimum intervention size for targeting amenable, resistant and randomly is given by the unstable fixed point corresponding to Eq. \ref{sieq11} shown in \cref{fig:theory}F.

%\generalTips
    \clearpage
    \subsection*{Supplemental Figures}
\setcounter{figure}{0} % set figs to start at 1
\renewcommand{\thefigure}{S\arabic{figure}} % start figures with "S"

% See 4maintextFigs.txt for functions and usage to help streamline making figures

\generateFig{theory}{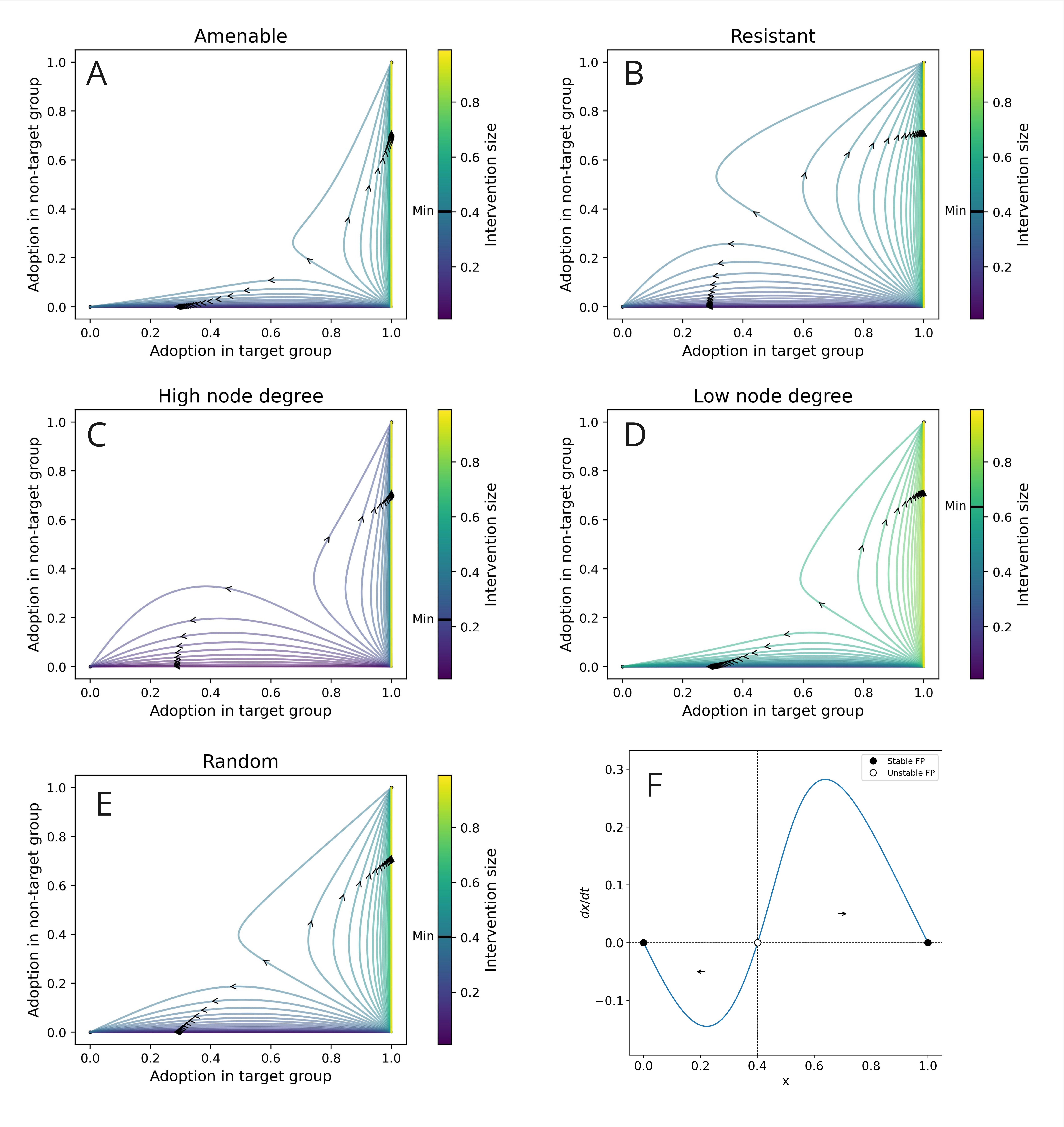}{0.95}
     {Adoption trajectories from theory}
     {Using a Markov Chain analysis (described in supplemental text), the trajectory of adoption is plotted for 5 different targeting strategies: amenable (A), resistant(B), high-degree nodes (C), low-degree nodes(D) and random (E). The fraction of adoption within both target and non-target group is plotted and the minimum intervention sizes are identified which are marked on the respective colorbars. The preference and degree distributions considered are the same as Fig 3. In panel F the fixed points corresponding to the Eq.S11 are plotted and we see that unstable fixed point $x^* \approx 0.4$ equals $\phi_{min}$ in panels A, B and E.    }
\theory

\generateFig{figsib}{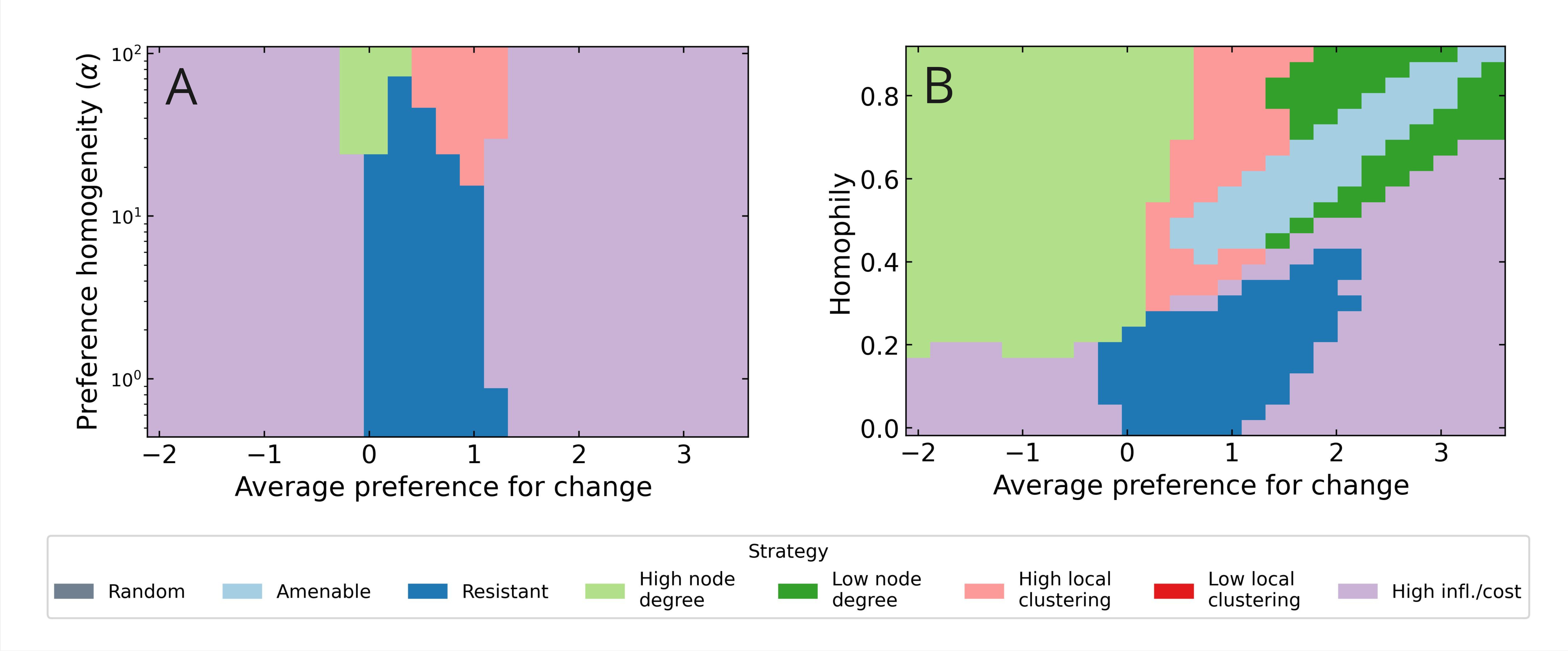}{1.0}
     {Targeting based on information about preferences as well the network}
     {We test a strategy to target individuals who have a high influence in the network relative to the cost of incentivizing them to change their behavior. This strategy is included in the analysis shown in figure 1 and 2 in the main text. The strategy outperforms other strategies for many different preference distributions (A) but its efficiency is negatively affected by segregation and clustering of preferences.}
\figsib

\generateFig{robust}{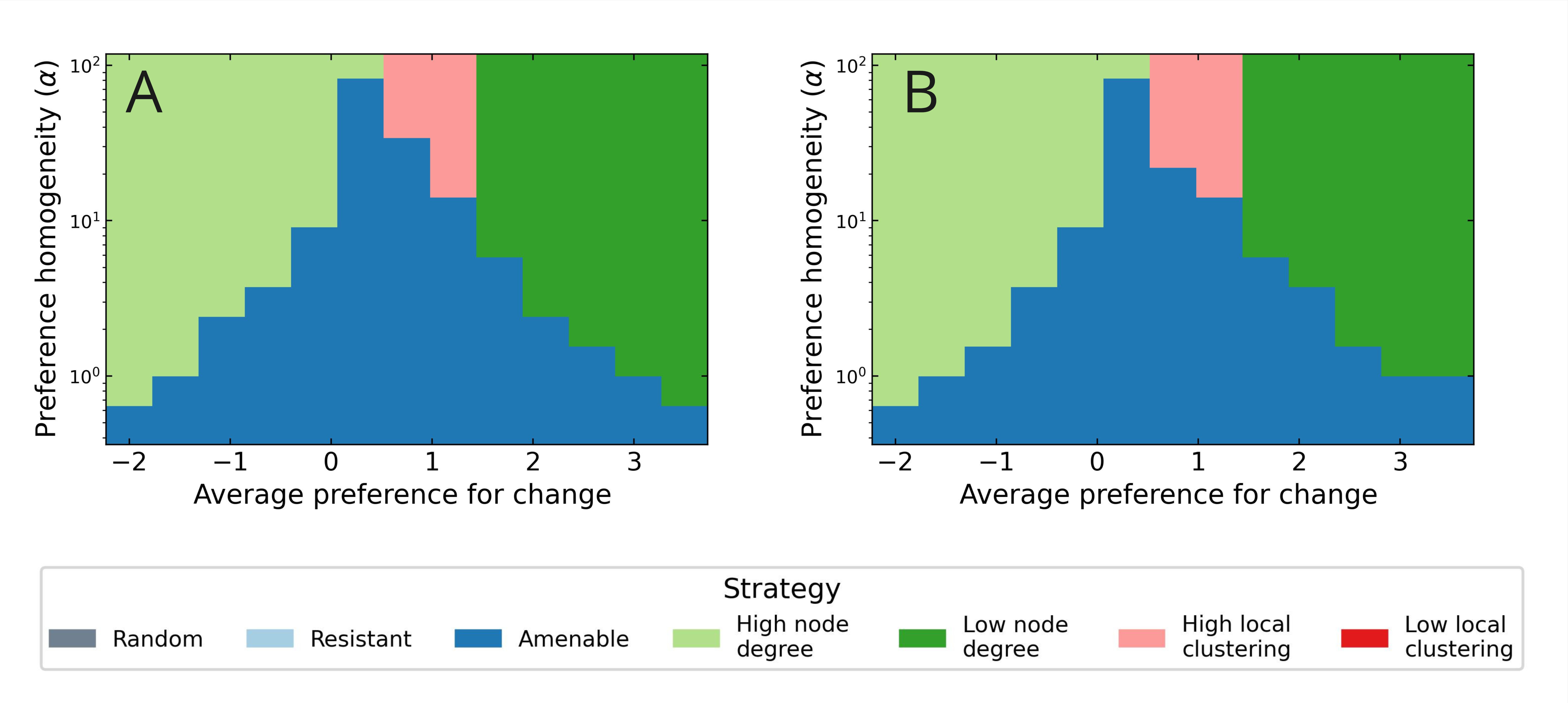}{1.0}
     {Robustness of analysis to target adoption level }
     {The cost-effective strategies across different preference distributions are plotted for achieving 85\% (A) and 95 \% adoption as compared to 90\% adoption which has been considered in the analysis presented in the paper.   }
\robust

\generateFig{figsig}{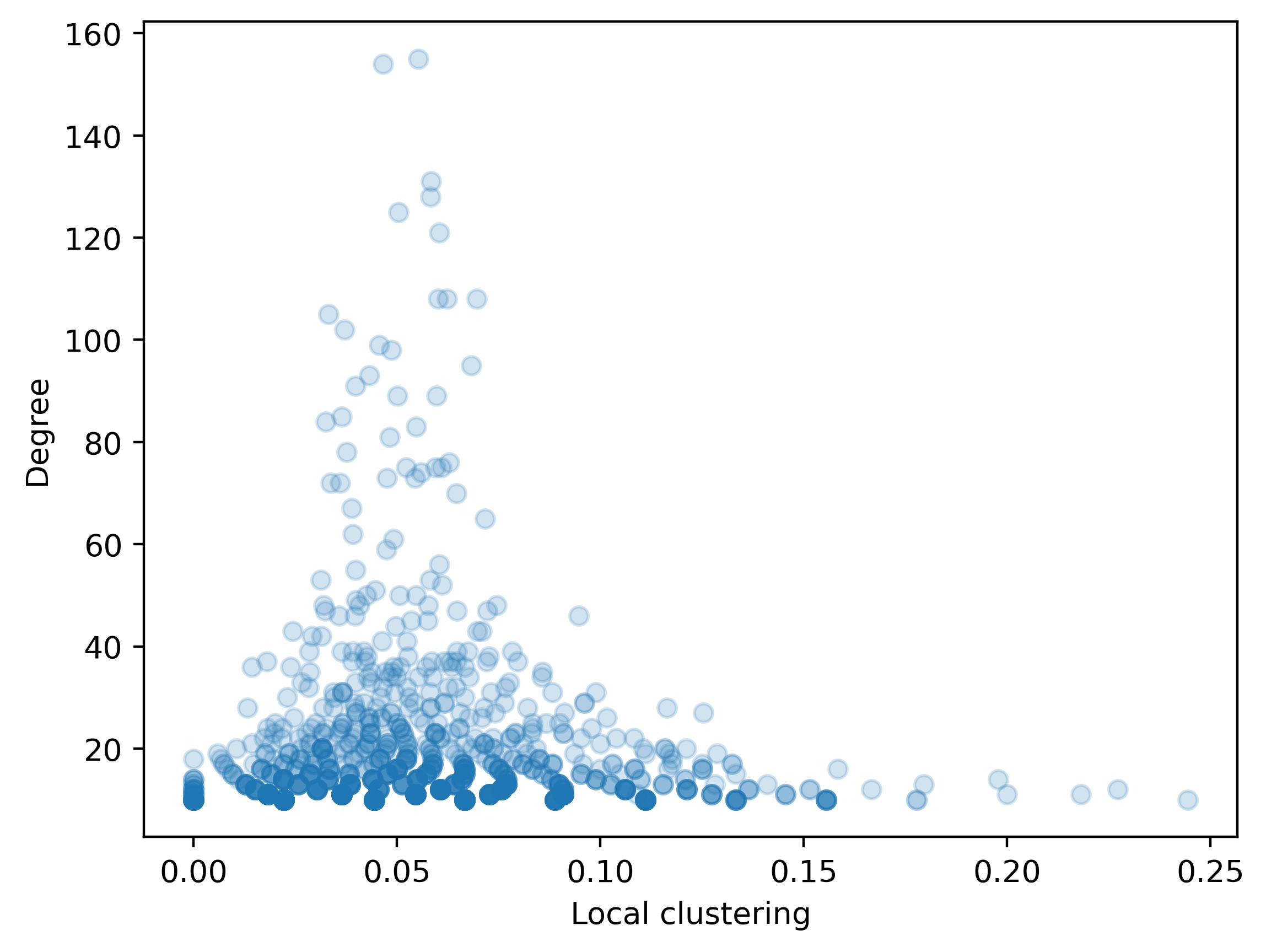}{0.8}
     {Node degree plotted against local clustering}
     {The node degree is plotted against the local clustering for a BA network with each point representing a node. Nodes with high local clustering are close to the network's periphery while nodes with low local clustering can be close to the center. This helps us understand the composition of target groups based on local clustering. }
\figsig

\generateFig{figsif}{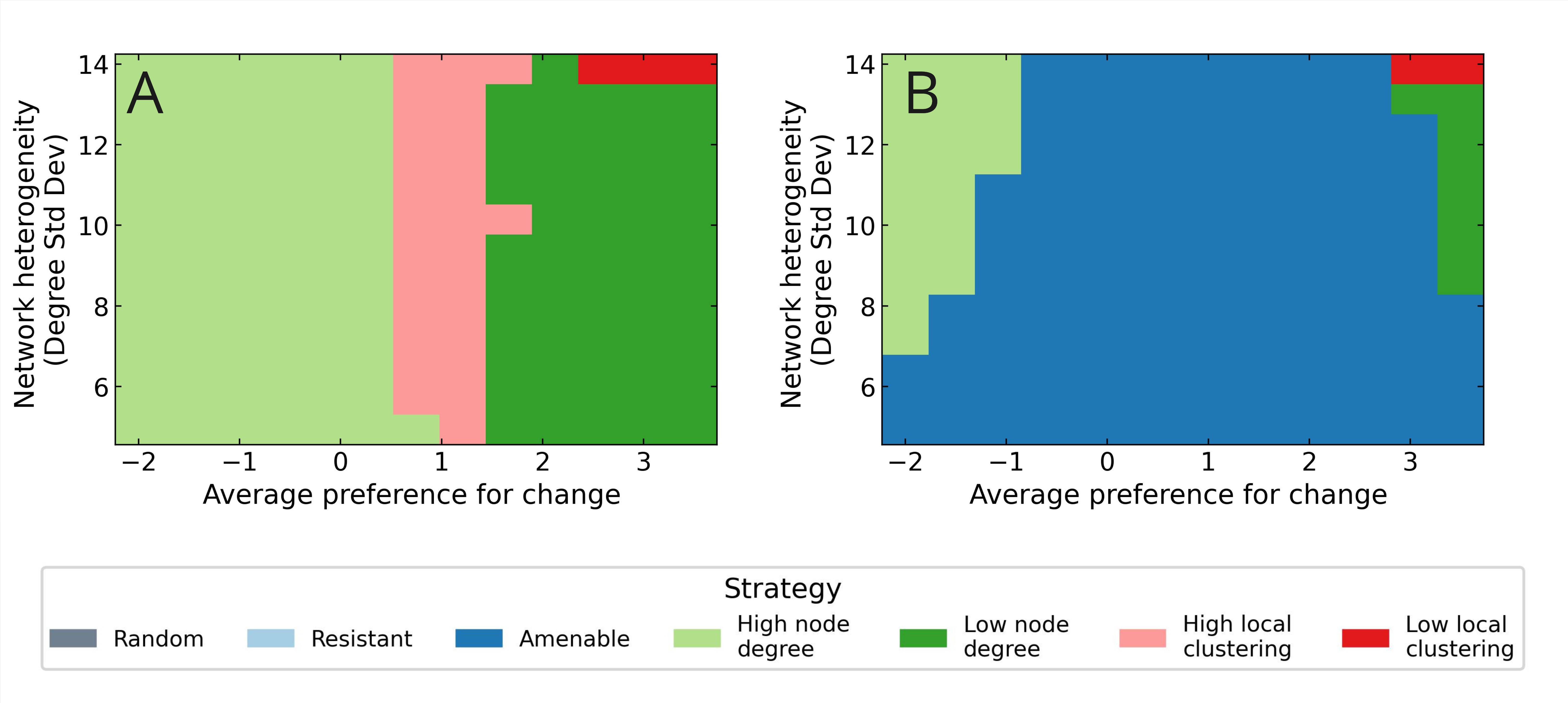}{1.0}
     {Varying network heterogeneity}
     {The strategies are tested for varying levels of network heterogeneity, with degree distribution varying from Poisson distribution to a scale-free distribution. The order of network-based strategies (A) is unaffected by network heterogeneity. For low network heterogeneity, the amenable strategy outperforms the network-based properties (B). The population size is 1000 with the average node degree for all networks = 20, alpha=2 and social influence parameter $\omega= 0.5$ }
\figsif
\generateFig{figsih}{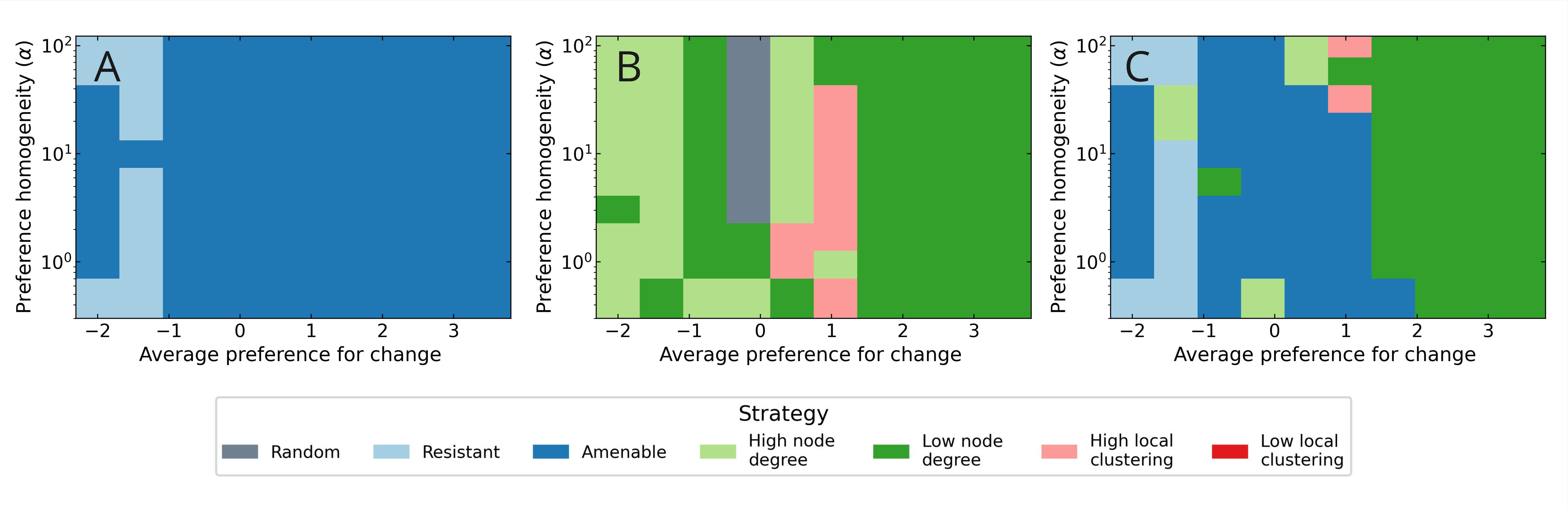}{1.0}
     {Simulation results for empirical networks}
     {The most cost-effective strategies to achieve 90 \% adoption are plotted for different preference distributions given by transformed Beta distribution, \(\text{Beta}(\alpha, \alpha)\), symmetrically centered around the average preference strength.  The average preference for change is varied and homogeneity is controlled through $\alpha$. The simulations are done on 74 networks from Add Health Dataset\supercite{jeon2015us}.  These networks vary in population size and average degree.  We test strategies based on information about preference distribution (A), networks (B), and the random targeting strategy. We then compare all strategies in C. The strategy which is cost-effective for the most number of networks is selected is plotted for different preference distributions.   We recover similar results obtained for synthetic Barabasi-Albert networks shown in Figure 1 in the main text. The social influence parameter $\omega =0.5$ .}
\figsih

\generateFig{figsie}{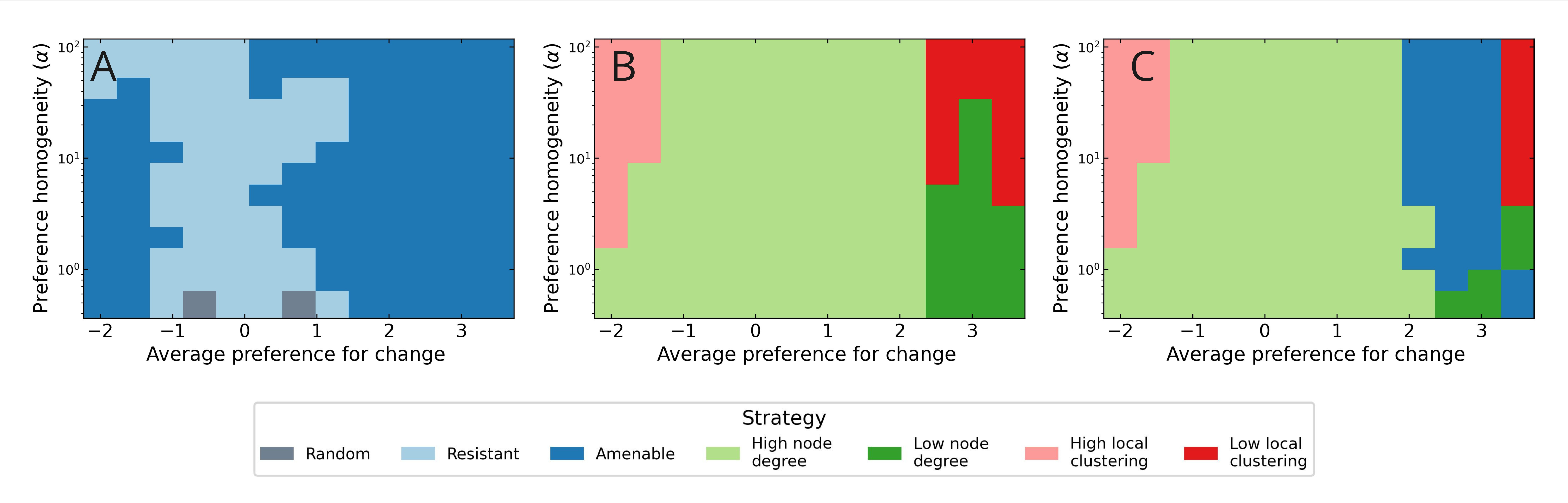}{1.0}
     {Model with ratcheted decision-making}
     {The most cost-effective strategies to achieve 90 \% adoption are plotted for different preference distributions given by transformed Beta distribution, \(\text{Beta}(\alpha, \alpha)\), symmetrically centered around the average preference strength.  The average preference for change is varied and homogeneity is controlled through $\alpha$. The model tested here is ratcheted, --- agents do not revert after adopting the new behavior.  We test strategies based on information about preference distribution (A), networks (B), and the random targeting strategy. We then compare all strategies in C. Compared to Figure 1 in the main text, the strategy targeting resistant individuals becomes cost-effective in parts of the parameter space (A). In B and C, targeting the highly connected nodes is cost-effective over a large parameter space as they sustain change in behavior without social pushback affecting them. The population size is 1000 and is connected via a Barabási-Albert network (min $k$=10) with social influence parameter $\omega =0.5$.}
\figsie

\generateFig{figsid}{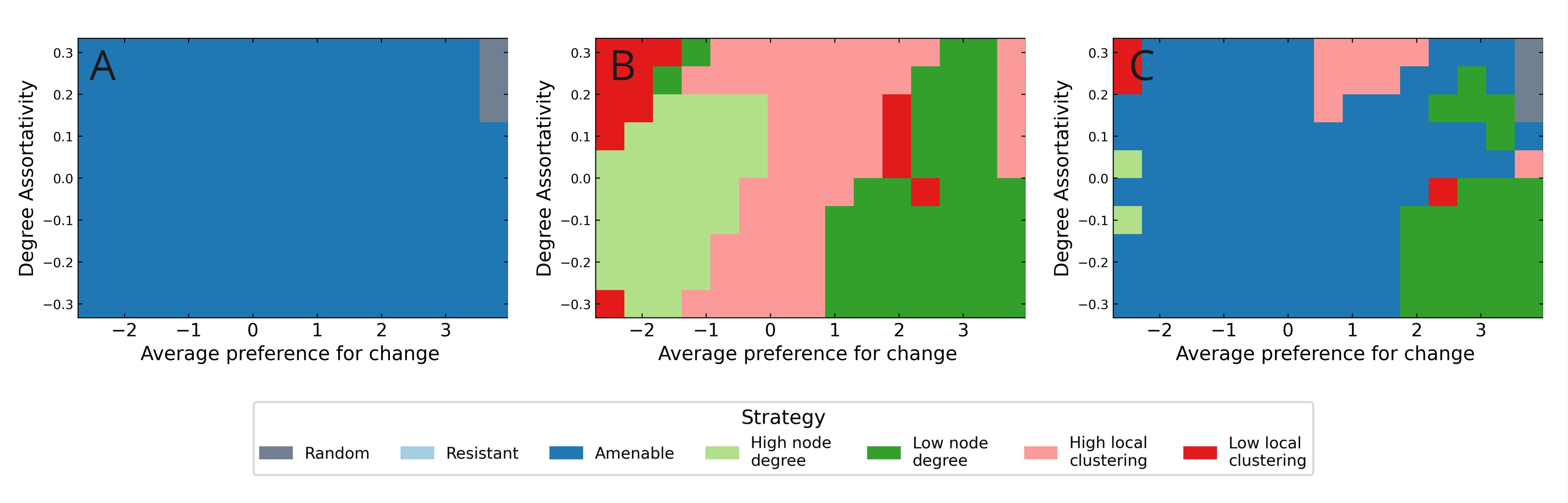}{1.0}
     {Effect of degree assortativity on the efficiency of network-based targeting}
     {The most cost-effective strategies to achieve 90 \% adoption are plotted for different preference distributions given by transformed Beta distribution, \(\text{Beta}(\alpha, \alpha)\), symmetrically centered around the average preference strength.  The average preference for change and degree assortativity is varied. We test strategies based on information about preference distribution (A), networks (B), and the random targeting strategy. We then compare all strategies in C.  In B, the degree-based strategies become relatively less effective as the assortativity increases. The population size is 1000 and is connected via Barabási-Albert networks (min $k$=10) rewired to tune degree assortativity with social influence parameter $\omega =0.5$.}
\figsid

\generateFig{figsic}{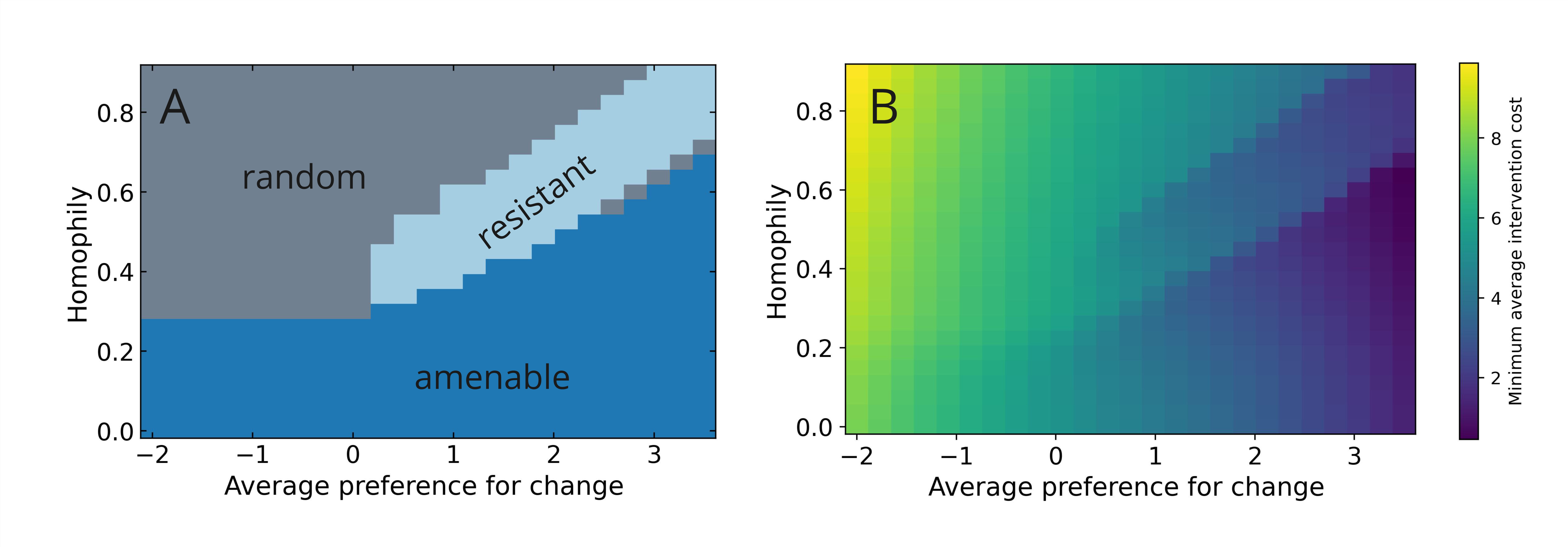}{1.0}
     {Segregation in more amenable populations can make preference-based targeting more cost-effective}
     {In B we show the average costs of incentives corresponding to the best-performing preference-based strategies shown in A (Fig2B in main text). For a high average preference for change, segregation of preferences decreases the cost of interventions. Clustering of targeted individuals prevents backsliding increasing the efficiency of strategies targeting resistant and amenable individuals. }
\figsic

\generateFig{figsia}{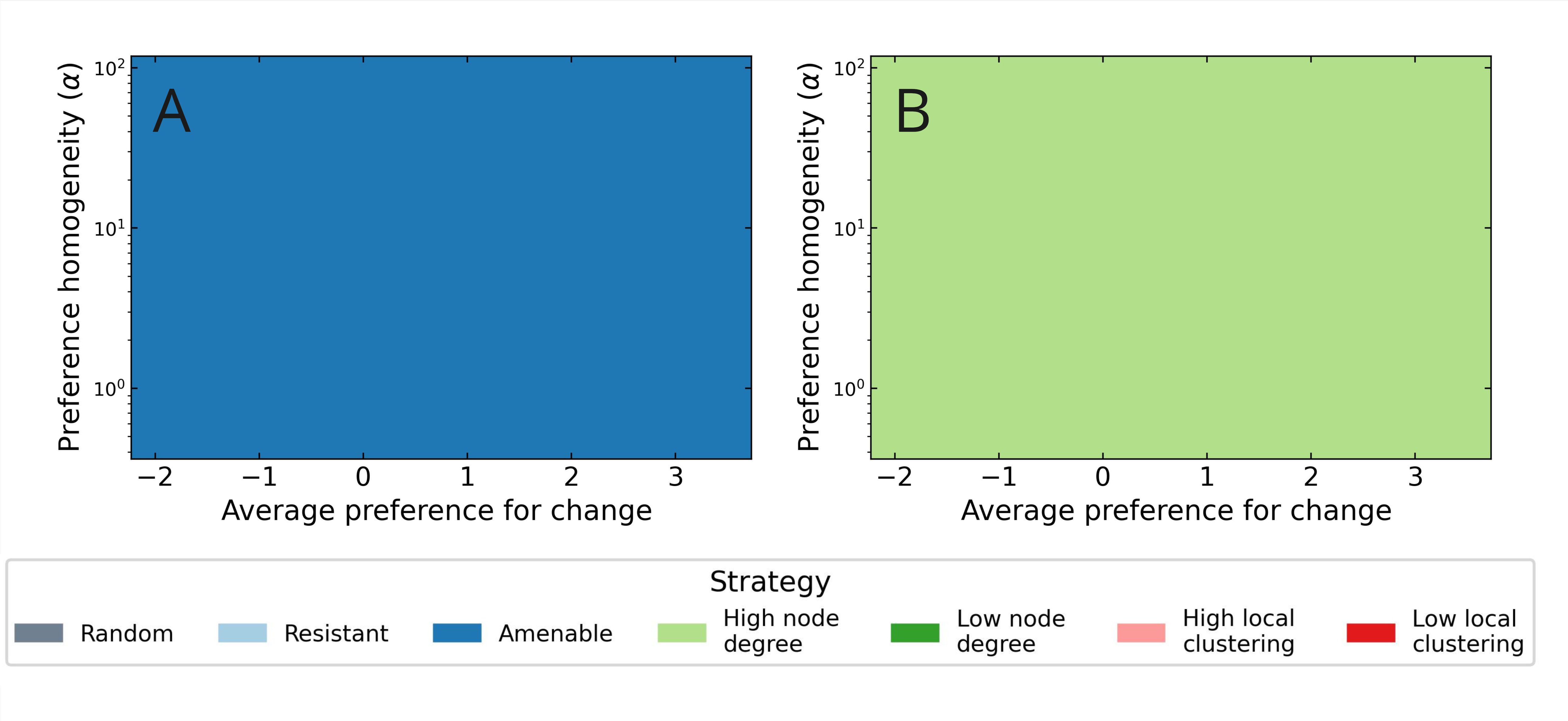}{1.0}
     {Model with social influence as a function of the fraction of adopters in the neighborhood  }
     {The most cost-effective strategies to achieve 90 \% adoption are plotted for different preference distributions given by transformed Beta distribution, \(\text{Beta}(\alpha, \alpha)\), symmetrically centered around the average preference strength.  The average preference for change is varied and homogeneity is controlled through $\alpha$. We test strategies based on information about preference distribution in A, and compare all strategies in B. Among strategies based on preferences (A) the amenable strategy is cost-effective across all preference distributions. However, in B, targeting the highly connected nodes is cost-effective over the entire parameter space as a higher degree does not translate to greater social pressure.  The population size is 1000 and is connected via a Barabási-Albert network (min $k$=10) and the maximum social influence an individual receives is 10. }
\figsia

\generateFig{figsix}{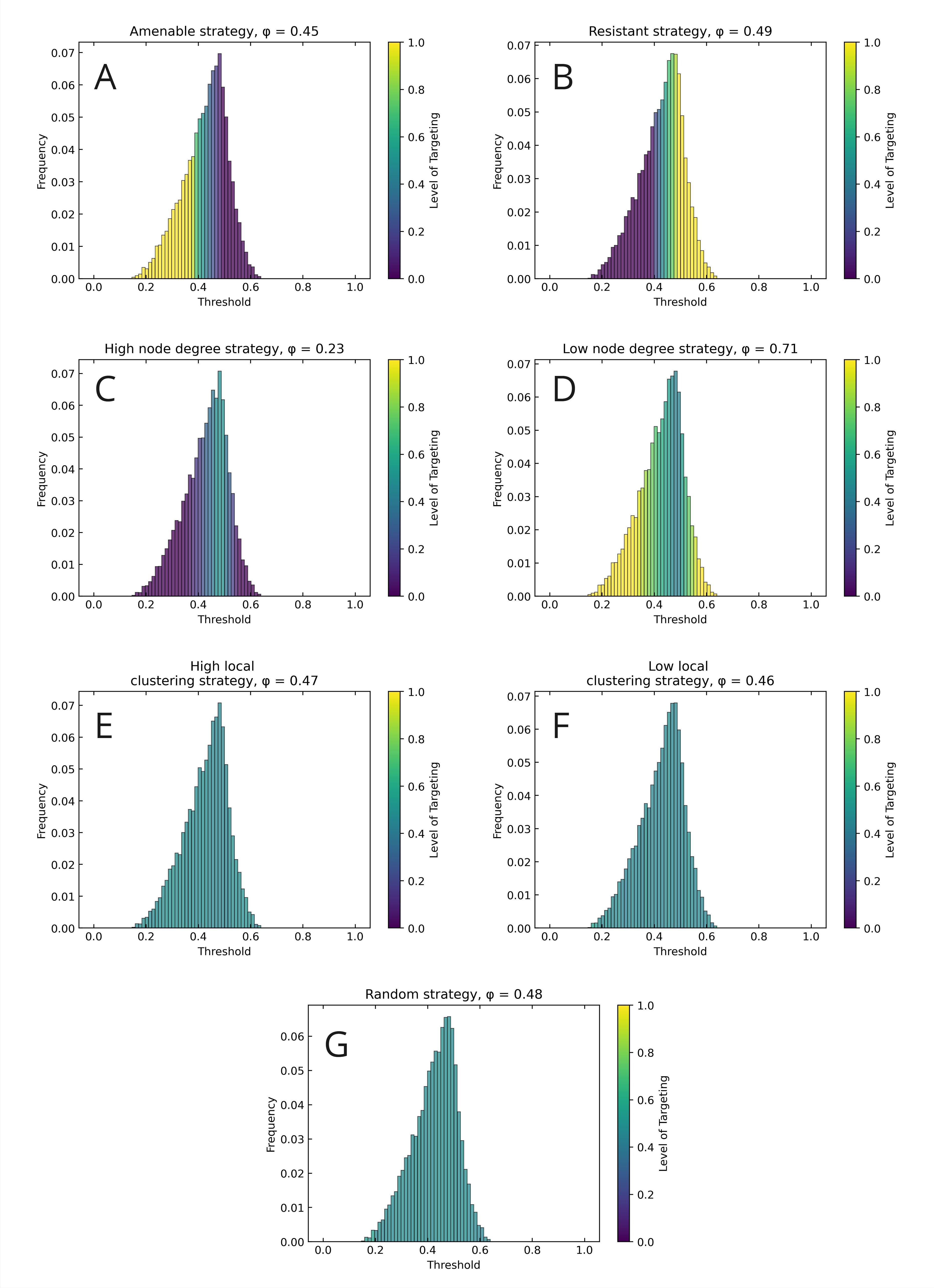}{1.0}
     {Mapping interventions to the threshold distribution}
     {Threshold is the fraction of adoption in the neighborhood that an agent needs before adopting the behavior. The threshold distributions are plotted for the population corresponding to Fig. 3. The spread of minimum interventions according to the different targeting strategies is shown with colors. }
\figsix

\clearpage
    %\subsection*{Supplemental Tables}
\setcounter{table}{0} % set tables to start at 1
\renewcommand{\thetable}{S\arabic{table}} % start tables with "S"

%%%% sampleTable %%%%

\generateTab{sampleTable}{supplement/suppl_tabs/sampleTable}{.55}
    {\textbf{\color{sectioncolor}\normalsize Example table.}} % Main caption, note the manual formatting included here. While tedious, this is the easiest way to format supplemental table titles. Just include the code before the title text in any new tables.
    {I use \texttt{Excel2LaTeX.xla} to create easy tables. There are also other online websites. Look for one that is booktabs enabled. Here's good advice about making simple readable tables: \url{https://users.umiacs.umd.edu/~jbg/images/tables.gif}} % Small caption
%\sampleTable

% Example of how to make a sideways table
% \generateSidewaysTab{sampleTableSideways}{supplement/suppl_tabs/sampleTable}{.3}
%     {\textbf{\color{sectioncolor}\normalsize Sideways Table.}} % Main caption
%     {Wow! This might be really useful for tables with lots of columns and fewer rows.} % Small caption
% \sampleTableSideways

\clearpage

\end{document}